\documentclass[a4paper,11pt]{article}
\pdfoutput=1 

\usepackage{jcappub} 
\usepackage{xcolor,bm}

\newcommand{\hkpc}{\ h^{-1}\text{kpc}}
\newcommand{\hMpc}{\ h^{-1}\text{Mpc}}

\newcommand{\ihMpc}{\ h\text{Mpc}^{-1}}

\newcommand{\be}{\begin{equation}}
\newcommand{\ee}{\end{equation}}

\newcommand{\kms}{{\rm km~s^{-1}}}

\newcommand{\vx}{\mathbf{x}}

\newcommand{\lya}{Ly$\alpha$}

\newcommand{\lyaf}{Ly$\alpha$ forest}
\newcommand{\lyax}{Ly$\alpha$ $\times$ halo}

\newcommand{\PoD}{$P_{\rm 1D}$}

\usepackage{csvsimple}
\usepackage{graphicx}
\usepackage{subcaption}
\usepackage{float}
\usepackage[noabbrev]{cleveref}
\newcommand{\edits}[1]{{\textcolor{red}{#1}}}

\title{\boldmath Non-linearities in the Lyman-$\alpha$ forest and in its cross-correlation with dark matter halos}

\author[a,b]{Jahmour J. Givans}
\author[c,d]{, Andreu Font-Ribera}
\author[e]{, An\v{z}e Slosar}
\author[f,d]{, Louise Seeyave}
\author[g,b,d]{, Christian Pedersen}
\author[h]{, Keir K. Rogers}
\author[i]{, Mathias Garny}
\author[j,c]{, Diego Blas}
\author[k,l]{and Vid Ir\v{s}i\v{c}}

\affiliation[a]{Department of Astrophysical Sciences, Princeton University, 4 Ivy Lane, Princeton, NJ 08540, USA}
\affiliation[b]{Center for Computational Astrophysics, Flatiron Institute, 162 5th Ave, New York, NY 10010, USA}
\affiliation[c]{Institut de F\'{i}sica d'Altes Energies (IFAE), The Barcelona Institute of Science and Technology, 08193 Bellaterra (Barcelona), Spain}
\affiliation[d]{Department of Physics and Astronomy, University College London, Gower Street, London WC1E 6BT, UK}
\affiliation[e]{Physics Department, Brookhaven National Laboratory, Upton, NY 11973, USA}
\affiliation[f]{Astronomy Centre, University of Sussex, Falmer, Brighton BN1 9QH, UK}
\affiliation[g]{Center for Cosmology and Particle Physics, Department of Physics, New York University, 726 Broadway, New York, NY 10003, USA}
\affiliation[h]{Dunlap Institute for Astronomy and Astrophysics, University of Toronto, 50 St. George Street, Toronto, ON M5S 3H4, Canada}
\affiliation[i]{Physik Department T31, Technische Universit\"at M\"unchen, James-Franck-Stra\ss{}e 1, D-85748 Garching, Germany
}
\affiliation[j]{Grup de F\'isica Te\`orica, Departament  de  F\'isica, Universitat  Aut\`onoma  de  Barcelona, Bellaterra, 08193 Barcelona, Spain}
\affiliation[k]{Kavli Institute for Cosmology, University of Cambridge, Madingley Road, Cambridge CB3 0HA, UK}
\affiliation[l]{Cavendish Laboratory, University of Cambridge, 19 J. J. Thomson Ave., Cambridge CB3 0HE, UK}

\emailAdd{jgivans@princeton.edu}
\emailAdd{afont@ifae.es}

\abstract{
Three-dimensional correlations of the Lyman-$\alpha$ (\lya) forest and cross correlations between the \lyaf\ and quasars have been
measured on large scales, allowing a precise measurement of the baryon acoustic oscillation (BAO) feature at redshifts $z>2$. 
These 3D correlations are often modelled using linear perturbation theory, 
but full-shape analyses to extract cosmological information beyond BAO will
require more realistic models capable of describing non-linearities present at smaller scales.
We present a measurement of the \lyaf\ flux power spectrum from large
hydrodynamic simulations --- the Sherwood simulations --- and compare it to different
models describing the small-scale deviations from linear theory.
We confirm that the model presented in Arinyo-i-Prats et al. (2015) fits the measured
3D power up to $k=10 \ihMpc$ with an accuracy better than 5\%, and show that the same
model can also describe the 1D correlations with similar precision.
We also present, for the first time, an equivalent study for the
cross-power spectrum of halos with the \lya\ forest, and we discuss different challenges
we face when modelling the cross-power spectrum beyond linear scales.
We make all our measured power spectra public in \url{https://github.com/andreufont/sherwood_p3d}.
This study is a step towards joint analyses of 1D and 3D flux correlations,
and towards using the quasar-\lya\ cross-correlation beyond BAO analyses.
}

\begin{document}
\maketitle

\flushbottom

\section{Introduction}
\label{sec:int}

The Lyman-$\alpha$ forest (\lyaf) is a series of absorption features in the
spectra of high-redshift quasars caused by the presence of
intervening neutral hydrogen.
During the last decade, the Baryon Oscillation Spectroscopic Survey (BOSS, \cite{2013AJ....145...10D}) and its extension
eBOSS \cite{2016AJ....151...44D} have obtained over 341,000 quasar spectra at $z>1.77$.
These surveys were able to measure the 3D autocorrelation of the \lyaf\ of quasars at $z>2.1$ 
\cite{2011JCAP...09..001S,2013A&A...552A..96B,2013JCAP...04..026S,2015A&A...574A..59D,2017A&A...603A..12B,2019A&A...629A..85D,2020ApJ...901..153D} and its 3D cross-correlation with the quasar positions \cite{2013JCAP...05..018F,2014JCAP...05..027F,2017A&A...608A.130D,2019A&A...629A..86B,2020ApJ...901..153D}, allowing researchers to detect baryon acoustic oscillations (BAO) around $z=2.3$ in 
increasingly larger datasets.

By examining the BAO scale as a function of redshift, researchers may measure both the angular diameter distance and the Hubble parameter across cosmic time. 
Recently, \cite{2021MNRAS.506.5439C} suggested that these measurements
can be improved through the inclusion of Alcock-Paczy\'{n}ski measurements \cite{1979Natur.281..358A}  from the full-shape (``broadband'') correlations. This information provides valuable insights into the nature of dark energy. 
Reference \cite{2021MNRAS.506.5439C} also suggested that a joint, broadband analysis of both power spectra could be used to measure the amplitude of redshift-space distortions (RSD) in the quasar sample.
RSDs are often parameterized with $f\sigma_8$, a quantity sensitive to the growth of structure.
Comparing growth of structure and expansion history allows us to test general relativity,
and the statistical power of these tests will depend on the minimum separation that we can
reliably model.
The first main goal of this publication is to quantify the deviations from linear theory
of both correlations (\lya\ auto-correlation and its cross-correlation with quasars),
and to compare models that would allow us to extend these studies to mildly non-linear scales.

The same quasar spectra can be used to study the clustering of matter on 
scales of a few megaparsecs
via measurements of the one-dimensional flux power spectrum, or \PoD\
\cite{Croft1998,Croft1999,McDonald2000,Gnedin2002,Croft2002,2004MNRAS.354..684V,McDonald2005,PD2013,Chabanier2019}.
In combination with the large-scale measurements from the cosmic microwave background,
these \PoD\ measurements allow tight constraints on the sum of the neutrino masses,
the shape of the primordial power spectrum
(both are referenced in \cite{Phillips2001,Verde2003,Spergel2003,Viel2004b,Seljak2005,Seljak2006,Bird2011,PD2015,PD2015b,PD2020}),
and several dark matter models
\cite{2006PhRvL..97s1303S,2013PhRvD..88d3502V,2017PhRvD..96b3522I,2017PhRvL.119c1302I,2017JCAP...12..013B,2018PhRvD..98h3540M,2019PhRvL.123g1102M,2019MNRAS.482.3227N,PD2020,2021PhRvL.126g1302R,2021PhRvD.103d3526R,Rogers:2021byl}.
However, while BAO analyses model the large-scale correlations using mainly linear
perturbation theory 
\footnote{The non-linear broadening of the BAO peak is usually based on Lagrangian 
perturbation theory, but it has a small effect on these high-redshift BAO results.},
\PoD\ analyses have to rely on hydrodynamic simulations to model non-linearities in the
distribution of gas and the physics of the intergalactic medium (IGM). 
So far, these BAO and \PoD\ analyses have been carried out as independent measurements.
Recently, \cite{2018JCAP...01..003F} suggested an algorithm to consistently
measure all relevant 1D and 3D correlations in the \lyaf. 
This joint analysis would not only improve the statistical uncertainty on some
cosmological parameters, but it would also make both measurements more robust against
possible systematic errors.
For instance, both measurements need to marginalise over the impact of 
Damped Lyman-$\alpha$ systems (DLAs), residual contamination from metal absorbers, 
and errors in the continuum fitting of the quasar spectra.
In order to carry out these joint analyses, however, we need a theoretical
framework to coherently model all scales involved.
The second main goal of this publication is to study whether models that describe the 3D
correlations in the \lyaf\ can consistently describe its 1D correlations on small scales.

Modeling the \lyaf\ 3D auto-power spectrum at non-linear scales has historically been done through a combination of perturbation theory and simulations. 
Generally, one begins by modeling the \lyaf\ flux fluctuation field at linear order using Kaiser's relation for the redshift-space clustering of matter tracers \cite{1987MNRAS.227....1K}. 
From there, one may proceed down a ``pure perturbation theory'' path or take a ``fitting function'' path. 
To proceed down the former, one adds higher-order terms consistent with symmetries of the forest to the linear flux fluctuation field \cite{2018JCAP...09..011G,2021JCAP...03..049G,2020PhRvD.102b3515G,2021JCAP...05..053C} and autocorrelates the result. 
Bias coefficients are found by fitting the model to simulation outputs. 
An advantage of the perturbation theory approach is that each term has a clear physical interpretation; a disadvantage is that these models break down on scales where perturbations approach order unity. 
The fitting function path was pioneered by \cite{2003ApJ...585...34M} and later improved in \cite{2015JCAP...12..017A}.
In this approach, the linear theory power spectrum is multiplied by a non-linear correction term with multiple free parameters. 
Just as before, the power spectrum model is fit to simulation outputs to find the optimal parameter values.
The fitting function approach typically provides better fits relative to the perturbation theory approach at higher wavenumbers (i.e. on non-linear scales), but a physical interpretation of the parameters is not always clear. 

In order to study different approaches to model the non-linearities in the \lyaf\ and its cross-correlation
with halos, it is useful to look at clustering measurements from hydrodynamic simulations.
Ideally these simulations would be large enough to contain many massive halos and have enough linear
modes to study the deviations from linear theory.
However, these hydrodynamic simulations need a high resolution in order to fully resolve the \lyaf,
and one needs to trade resolution and box size.
In this publication we use the outputs from some of the largest boxes in the Sherwood suite of
hydrodynamic simulations \cite{2017MNRAS.464..897B}, and present the most detailed measurements
of the 3D power spectrum of the \lyaf\ and of its cross-correlation with halos.
These allow us to compare different models for the \lyaf\ auto-correlation, and we present the first
detailed discussion on the non-linearities in its cross-correlation with halos.
\footnote{The cross-correlation function (in configuration space) was measured 
from \texttt{LyMAS} simulations in \cite{2016MNRAS.461.4353L}, but the authors did not
discuss extensions of linear theory.}

We organize this paper in the following way: we start in \cref{sec:sim} by briefly describing the Sherwood
simulations and data products that we use. 
In \cref{sec:analysis} we detail the methodology used to compare the 
different auto- and cross-power spectra measurements with models.
In \cref{sec:flux} we review existing models for the \lyaf\ power spectrum and
compare them to the measured one.
We do the same for the halo power spectrum in \cref{sec:halo}.
In \cref{sec:cross} we discuss models for the \lyax\ spectrum, and study its dependence with halo mass. 
We discuss the results and mention possible extensions in \cref{sec:discussion} before concluding in \cref{sec:conclusions}.
\section{Simulations}
\label{sec:sim}

We present a brief introduction to the Sherwood suite of hydrodynamic simulations
in \cref{sec:sim_details}, before describing the extraction of \lyaf\ data cubes
in \cref{sec:lya_grids} and grids of halo counts in \cref{sec:halo_grids}. 
We also provide measurements of auto- and cross-power spectra. 

\begin{figure}
    \centering
    \includegraphics[width=0.49\textwidth]{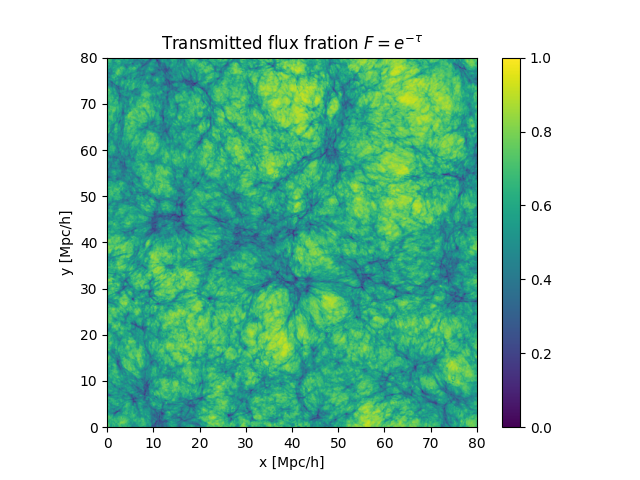}
    \includegraphics[width=0.49\textwidth]{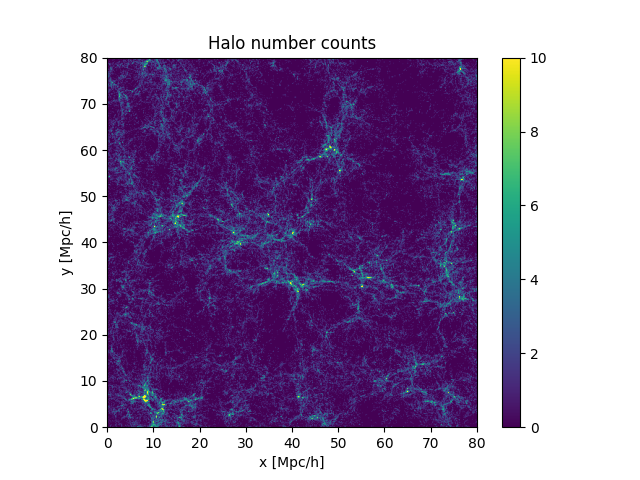}
    \caption{2D slab of the 3D grids extracted from the $z=2.8$ snapshot in the
    \texttt{L80\_N1024} simulation, averaged over a depth of $20 \hMpc$ 
    along the line-of-sight (perpendicular to the plot). 
    The \lyaf\ transmitted flux fraction (left) is strongly anti-correlated
    with the halo number counts (right).
    }
    \label{fig:2d_slabs}
\end{figure}

\subsection{The Sherwood simulations}\label{sec:sim_details}

The Sherwood simulation suite is a set of large, high-resolution hydrodynamic simulations of the intergalactic medium (IGM) with up to 17.2 billion particles in comoving volumes $10^3-160^3 \,h^{-3}$Mpc$^3$ \cite{2017MNRAS.464..897B}. 
The combination of box size and resolution makes them very well-suited for studies of the \lya\ forest in the mildly non-linear regime. 

All simulations were run using a modified version of the smoothed particle hydrodynamics code \texttt{P-Gadget3}, an updated version of \texttt{Gadget-2} \cite{2005MNRAS.364.1105S}. The initial conditions were generated at $z=99$ using the \texttt{N-GenIC} code \cite{2005MNRAS.364.1105S} according to an input power spectrum corresponding to a cosmology based on the best-fit flat $\Lambda$CDM model from the first results of the \textit{Planck} satellite \cite{2014A&A...571A..16P}, i.e. with $\Omega_m=0.308$, $\Omega_b=0.0482$, $h=0.678$, $\sigma_8=0.829$ and $n_s=0.961$.

A homogeneous ionising background is used to calculate photoionization and photo-heating rates assuming the gas is optically thin and in photoionization equilibrium. In this publication we will limit our study to snapshots in the relatively low-redshift range $2 < z < 3.2$ where BOSS, eBOSS and DESI have measured 3D correlations. As described in \cite{2017MNRAS.464..897B}, \lyaf\ measurements at these redshifts are less affected by the reionization model assumed in the simulation relative to measurements at higher redshifts. 

As is common in large simulations of the \lyaf, the computing time is dramatically reduced by removing gas particles at overdensities above $\Delta > 1000$ and temperatures below $T < 10^5 K$ and converting them to collisionless particles \cite{2004MNRAS.354..684V}.
This approximation, usually known as \texttt{Quick-Lya}, is valid on the low and intermediate densities that dominate the information in the \lyaf. 
However, as discussed in \cite{2017MNRAS.468.1893M,2019MNRAS.489.5381M}, this approximation
effectively removes all gas within the virial radius of halos. 

In order to properly simulate these high-density regions around massive halos,
it would not be enough to keep the gas particles removed by \texttt{Quick-Lya}.
We also expect these regions to be affected by different feedback mechanisms not
present in the simulations, including the ionizing radiation from quasars themselves. 
Moreover, we expect the majority of large neutral hydrogen absorbers, including 
Lyman limit systems and Damped \lya\ systems, to come from the vicinity of massive halos; 
the Sherwood simulations used in this analysis are optically thin, and do not incorporate
self-shielding corrections necessary to reproduce the abundances of these systems.
In the next sections we will have to keep this in mind when studying the \lyax\
cross-correlations on small scales. 

The main results discussed in this paper will come from the simulation
\texttt{L160\_N2048}, where the box size is $L=160 \hMpc$ and the
number of CDM particles and gas particles in the simulation is $N=2048^3$.
We present some results from the \texttt{L80\_N2048} and the
\texttt{L80\_N1024} simulations in \cref{app:alt}. 
Even though the Sherwood simulations are the state of the art in terms of hydrodynamical simulations of the \lyaf, they have not formally converged, and this should be taken into account when interpreting the results.

\lya\ datasets typically cover a wide redshift range $2 < z < 5$.
While \lya\ BAO results have an effective redshift around $z=2.3$ 
\cite{2020ApJ...901..153D}, \PoD\ analyses have an effective redshift closer
to $z=3$ \cite{McDonald2006}.
Most of the results presented in this publication are from a snapshot taken at $z=2.8$, 
but in \cref{app:redshift} we show results at $z=2.0, 2.4, 2.8,$ and $3.2$.

\subsection{Simulated \lyaf\ data} \label{sec:lya_grids}

We extract \lya\ absorption skewers using the public software
\texttt{fake\_spectra}\footnote{\url{https://github.com/sbird/fake_spectra}.} 
which interpolates the gas properties into the requested lines of sight
and computes the \lya\ optical depth including redshift-space distortions
\cite{fakespec}.

The IGM around $z=2-4$ is smoothed by gas pressure, suppressing structure
on scales smaller than the filtering length $\lambda_F \sim 50 \hkpc$ 
\cite{2017ApJ...837..106O}.
Unfortunately, memory availability in our post-processing of the simulations
limits our analysis to a 3D grid of $(1024)^2 \times 2048$ cells, with a cell
size of $156 \hkpc$ in the transverse direction and $78 \hkpc$ along the
line of sight. 
In \cref{app:alt} we show that while the measured power spectra are
affected by the resolution of these grids, they mostly affect the very high-k
modes and do not change the main conclusions presented in this paper. 

The left panel of \cref{fig:2d_slabs} shows a 2D slab of the \lyaf\
transmitted flux fraction from the snapshot at $z=2.8$ of the smaller 
\texttt{L80\_N1024} simulation. This is displayed using a grid of $512^2 \times 1024$ to
match the resolution of the main grid used in this section.

\begin{figure}
    \centering
    \includegraphics[width=0.9\textwidth]{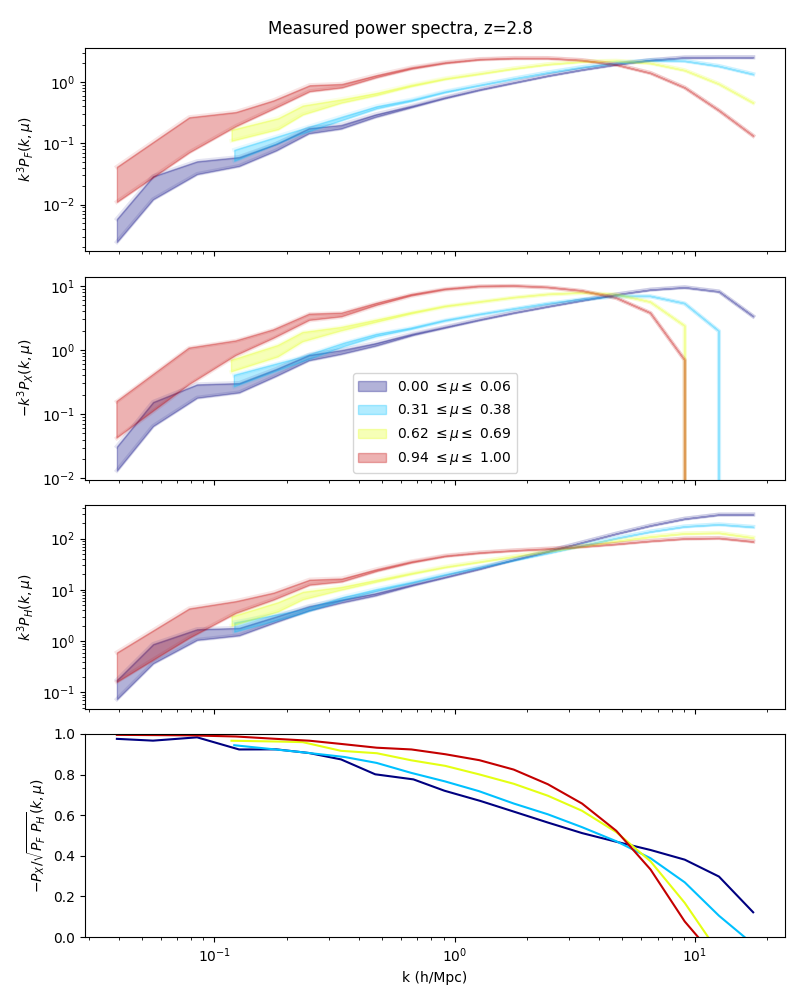}
    \caption{Measured 3D power spectra at $z=2.8$, as functions of
    wavenumber $k$, for the flux ($P_F$), halos ($P_H$, shot-noise subtracted)
    and their cross-correlation ($P_X$, multiplied by $-1$).
    The different colours show the power spectra in 4 of the 16 bins in $\mu$ (orientation with respect to the line of sight, with $\mu=1$ being parallel to the line of sight).
    The bottom panel shows the correlation coefficient. 
    The sharp cutoff at high wavenumber for the cross-power spectrum arises from
    a change of sign in the cross-power spectrum.}
    \label{fig:measured_p3d}
\end{figure}

All \lya\ statistics in this paper are computed for fluctuations around a mean
transmitted flux fraction, defined as:
\begin{equation}
    \delta_F(\vx) \equiv \frac{F(\vx)}{\bar F} - 1 ~,
\end{equation}
where $F(\vx)=e^{-\tau(\vx)}$ is the transmitted flux fraction,
and $\tau(\vx)$ is \lya\ optical depth in redshift space. 
The upper panel of \cref{fig:measured_p3d} shows the measured power
spectrum of the \lya\ fluctuations at $z=2.8$ in the larger box
(\texttt{L160\_N2048}) as a function
of wavenumber $k$ and the cosine of the angle between the Fourier mode and the
line of sight $\mu$ (modes with $\mu=1$ are parallel to the line of sight). 

The power spectra are computed using the average of the square of the norm for all
Fourier modes in the box, binned in 16 bins in $\mu$ and 20 log-spaced bins
in wavenumber $k$, ranging from the fundamental mode of the box 
($k_0=2 \pi / L = 0.039 \ihMpc$) to a maximum
wavenumber of $20 \ihMpc$ which is half the Nyquist frequency. 

On large scales (small wavenumbers), one can see that the line-of-sight power (red)
is boosted by a constant factor with respect to the transverse power (dark blue),
as expected in the case of linear bias and linear redshift-space distortions
\cite{1987MNRAS.227....1K}. 
As discussed later in \cref{sec:flux}, non-linearities become important around
$k = 0.5 \ihMpc$.
Non-linear peculiar velocities and thermal broadening suppress the power along
the line of sight, before gas pressure smooths the power on all directions
around $k = 10.0 \ihMpc$.

In order to reduce the cosmic variance in our measurements, we extract \lya\
absorption skewers along each of the three axes of the box, using a different
component of the 3D velocity vector to include redshift-space distortions
\cite{2021MNRAS.500..259S}.
All power spectra discussed in this paper are computed as the average of the
power spectra computed along each of the three axes.
The errorbars shown in the plots are estimated from the number of Fourier modes
entering each band power; they ignore the correlation between measurements along the different axes and the correlation between different Fourier modes within a band power (or across band powers).

The power spectra measurements used in this paper are available online in
\url{https://github.com/andreufont/sherwood_p3d}.

\subsection{Halo grids}\label{sec:halo_grids}

The output of the Sherwood simulations includes a halo catalog constructed with
a \textit{friends of friends} algorithm. 
In order to measure the halo power spectrum and the \lyax\ spectrum,
we start by discretising this catalog in the same 3D grid that we used for 
the \lya\ skewers.
We do this using a nearest-grid-point interpolation.
The third panel of \cref{fig:measured_p3d} shows the measured halo power
spectrum once the shot-noise, estimated as the inverse of the halo density,
has been subtracted.

The second panel of the same figure shows the measured \lyax\ spectrum,
multiplied by $(-1$) to make it a positive quantity on the largest scales
(as seen in \cref{fig:2d_slabs}, these fields are anti-correlated).
On small, non-linear scales, the cross-power changes sign and becomes positive, 
causing a sharp feature in this logarithmic plot.
It is important to remember that the simulations used the \texttt{Quick-Lya}
approximation, replacing gas particles in very high-density regions with
collisionless particles, and resulting in a transmission field positively
correlated with the halo density on scales comparable to their virial radius.

Finally, the fourth panel of \cref{fig:measured_p3d} shows the correlation
coefficient between the two grids, defined as 
\begin{equation}
 \rho(k,\mu) = \frac{P_X(k,\mu)}{\sqrt{P_F(k,\mu) P_H(k,\mu)}} ~.
\end{equation}
On very large scales ($k < 0.3 \ihMpc$) the two fields are highly anti-correlated,
with $\rho < -0.9$.
Even on smaller scales of $k = 1 \ihMpc$, the line-of-sight (red) remain 90\%
anti-correlated, while the anti-correlation has dropped to 70\% for the
transverse modes (dark blue).
This strong anti-correlation motivates the model for the \lyax\ spectrum discussed
in \cref{sec:cross}.

\begin{figure}
    \centering
    \includegraphics[width=0.49\textwidth]{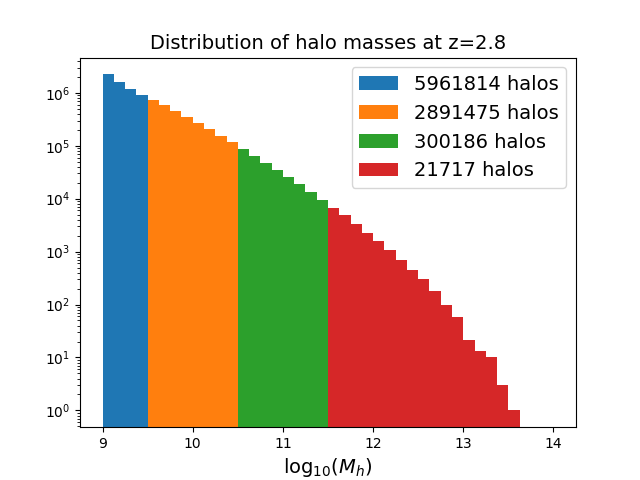}
    \includegraphics[width=0.49\textwidth]{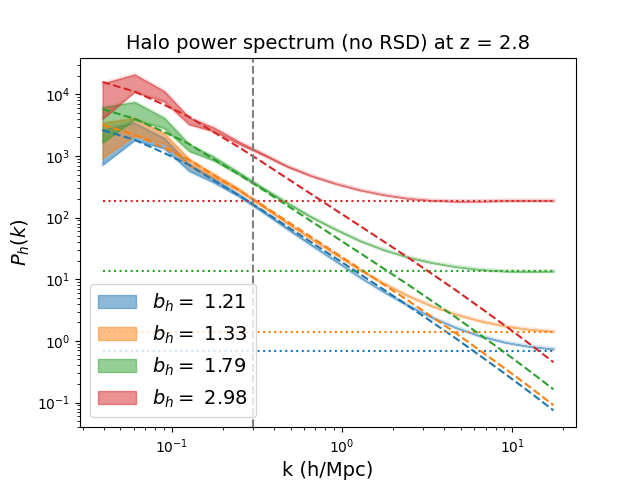}
    \caption{(left) Distribution of halo masses for the halo catalog at $z=2.8$.
    The different colors identify the four halo mass bins for which we present 
    its power spectrum (right). 
    Note that in this figure we have not added RSDs to the halos and we have not
    subtracted the Poissonian shot noise (dotted lines).
    Colored dashed lines show the best-fit one-loop model, as described
    in \cref{ss:halo_no_RSD}, with halo biases reported in the legend.
    The vertical gray dashed line shows the maximum wavenumber ($k = 0.3 \ihMpc$)
    used in the fit.
    }
    \label{fig:halo_masses}
\end{figure}

Our principal motivation in carrying out this study is to improve the 
modeling of the \lya\ $\times$ quasar spectrum; its correlation function analog was
recently measured by BOSS and eBOSS 
\cite{2013JCAP...05..018F,2014JCAP...05..027F,2017A&A...608A.130D,2019A&A...629A..86B,2020ApJ...901..153D}
and is one of the key measurements expected from DESI.
Quasars are highly biased tracers, with typical values of $b_q=3.5$ measured
at $z \sim 2.5$ \cite{2013JCAP...05..018F}. They are typically hosted
in dark matter halos larger than $\sim 10^{12} M_\odot$ at the
relevant redshift range.
Unfortunately, as can be seen in \cref{fig:halo_masses}, most of the halos in our
simulated box are much smaller --- we have only a few thousand halos with
a mass similar to those expected to host quasars. 
Therefore, most plots in this publication use all halos regardless of mass, 
except in \cref{sec:cross} where we also show the \lyax\ power for 
the highest mass bin. 

\section{Analysis method}
\label{sec:analysis}

In the next few sections we will present comparisons between different clustering models
and the measurements from the simulations presented in \cref{fig:measured_p3d}.
We will fit models by minimising a pseudo-$\chi^2$ function:
\begin{equation}
\label{eq:pseudo_chi2}
  \chi^2(\mathbf{\theta}) = \sum_i w_i \left[ P_i - P(k_i,\mu_i | \mathbf{\theta}) \right]^2~,
\end{equation}
where $P_i$ is the measured power spectrum, $P(k_i,\mu_i | \mathbf{\theta})$ is the
theoretical prediction for a set of parameters $\theta$, and $w_i$ is the weight given
to a particular band power.

We could have weighted each band power using the inverse of its uncertainty,
computed from the number ($N^{\rm 3D}_i$) of Fourier modes that contribute to each band.
However, in this study we combine the power spectrum measured along the
three axes of our box, and their modes are not independent. 
Moreover, as discussed in references \cite{2003ApJ...585...34M,2015JCAP...12..017A}, the standard
fit would be dominated by the high-k data points that have very small errorbars; that would make it difficult to model the large scales comparable to our box size.
Following \cite{2003ApJ...585...34M,2015JCAP...12..017A}, we modify the standard
weights by adding a \textit{noise floor} ($\epsilon$):
\begin{equation} \label{eq:weights}
  w_i^{-1/2} = P_i \left[ \frac{1}{\sqrt{N_i}} + \epsilon \right] ~.
\end{equation}

This noise floor is set by default to $\epsilon=0.05$ as in \cite{2015JCAP...12..017A},
and it is most relevant for the fits of the flux power spectrum. 
The shot-noise in the halos already provide a natural noise floor for the halo
auto power spectrum and its cross-power spectrum with the \lya\ forest.
We only include data points with $k<k_{\rm max}$. 
When fitting the flux power spectrum we use by default $k_{\rm max}=10 \ihMpc$,
and we discuss the impact of varying these settings in \cref{app:alt}.
The values of $k_{\rm max}$ used in the auto-correlation of halos and in the \lyax\ measurements are discussed in later sections.

\section{Modelling the \lyaf\ flux power spectrum}
\label{sec:flux}

The standard linear theory model for \lyaf\ flux fluctuations is
\begin{equation}\label{eq:linear_flux}
    \delta_F=b_F\delta+b_\eta \eta.
\end{equation}
Here, $\delta$ is the matter overdensity,  
\begin{equation}
    \eta=-\frac{1}{aH}\frac{\partial v_{||}}{\partial r_{||}}
\end{equation}
is the dimensionless gradient of the peculiar velocity $v_{||}$ along the line of sight (represented by comoving coordinate $r_{||})$, $a$ is the scale factor, and $H$ is the Hubble parameter. 
The fields in \cref{eq:linear_flux} are evaluated in redshift space since the \lyaf\ is inherently a redshift-space phenomenon.
However, at this order in perturbation theory there is no distinction between
real-space coordinate \textbf{r} and redshift-space coordinate \textbf{s}. 
Bias coefficients $b_F$ and $b_\eta$ are defined as the partial derivatives of $\delta_F$ with respect to $\delta$ and $\eta$, respectively.
These coefficients do not have the same interpretation as in the case for galaxies (see \cite{2015JCAP...12..017A} for further discussion).
The power spectrum associated with \cref{eq:linear_flux} is
\begin{equation}\label{eq:kaiser_flux}
    P_\textrm{F}(k,\mu)=b_F^2(1+\beta_F\mu^2)^2P_L(k),
\end{equation}
where $\beta_F=b_\eta f/b_F$ is the RSD parameter with $f$ representing the logarithmic growth rate, $\mu$ is the cosine of the angle between Fourier mode \textbf{k} and the line of sight, and $P_L$ is the linear matter power spectrum.
We refer to this model as the \textit{Kaiser} model \cite{1987MNRAS.227....1K}, 
but it is important to note that the definition of $\beta_F$ here includes an extra bias parameter $b_\eta$ that is not needed when studying the clustering of point sources.

\subsection{Multiplicative corrections to the linear model}

Two decades ago, \cite{2003ApJ...585...34M} presented a model for the flux
power spectrum that was able to fit the 3D power measured in hydro-PM simulations
down to very small scales ($k = 10 \ihMpc$).
It accomplishes this by introducing a multiplicative correction to the linear 
model, containing 8 free parameters describing non-linearities and the physics
of the IGM. 
That correction term is 
\begin{equation}\label{eq:D0}
    D_{\rm NL}(k,\mu) = \exp\left\{ \left[ \frac{k}{k_{\rm{NL}}} \right]^{\alpha_{\rm{NL}}} - \left[ \frac{k}{k_P} \right]^{\alpha_P} - \left[ \frac{k_{||}}{k_{V}(k)} \right]^{\alpha_V} \right\},
\end{equation}
with $k_V(k)=k_{v0}(1+k/k_{v1})^{a_{v1}}$. 
The first term in the exponential accounts for isotropic nonlinear growth, the second for isotropic suppression due to pressure, and the third for line-of-sight suppression arising from non-linear peculiar velocities and thermal broadening \cite{2003ApJ...585...34M}. 
The product of \cref{eq:linear_flux,eq:D0} is what we will refer to as the \textit{McDonald} model.  

A similar analysis was carried out the following decade on outputs from GADGET-II
simulations \cite{2015JCAP...12..017A}, and using a modified correction:
\begin{equation}\label{eq:D_NL}
    D_{\rm NL}(k,\mu) = \exp \left\{ \Big[q_1\Delta^2(k)+q_2\Delta^4(k)\Big] \left[1-\left(\frac{k}{k_v}\right)^{a_v}\mu^{b_v} \right]-\left(\frac{k}{k_p}\right)^2\right\},
\end{equation}
where
\begin{equation} \label{eq:delta2}
    \Delta^2(k) \equiv \frac{k^3}{2\pi^2}P_L(k)
\end{equation}
is the dimensionless linear matter power spectrum. 
When comparing this model to the measured flux power spectrum from \cref{sec:sim},
we find that the best-fit value for the second order non-linear growth ($q_2$)
is very close to zero.
Therefore we decided to ignore this extra term, reducing the number of free
parameters to 5 (compared to the 8 parameters in the McDonald model).
We will call the product of \cref{eq:linear_flux,eq:D_NL} the \textit{Arinyo} model. 

\begin{figure}
    \centering
    \includegraphics[width=0.95\textwidth]{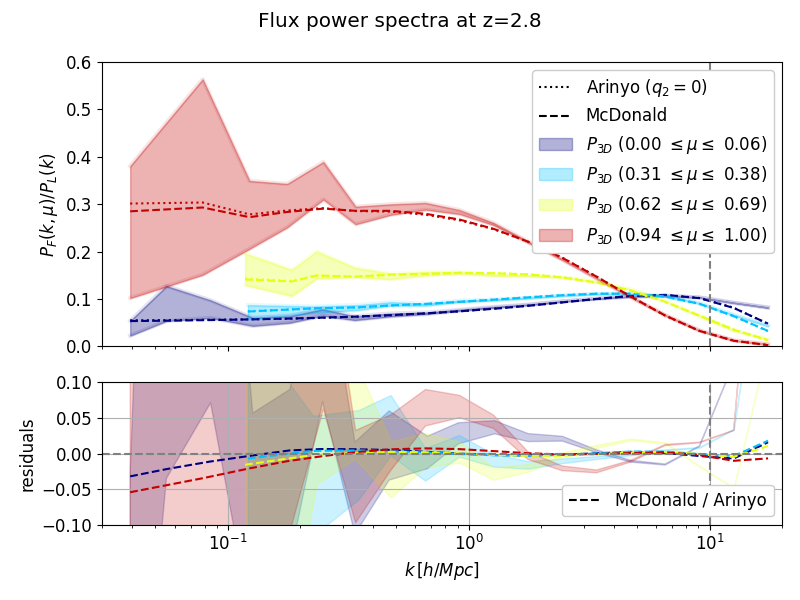}
    \caption{(Top) Model comparison between the Arinyo (with $q_2=0$)
    and McDonald models, and the measured power spectra at $z=2.8$. 
    The 3D power spectra are shown as functions of wavenumber $k$ and
    orientation $\mu$, divided by the linear power. 
    (Bottom) Residuals with respect to the Arinyo model.}
    \label{fig:ari_vs_mcd}
\end{figure}

The top panel of \cref{fig:ari_vs_mcd} shows the best-fit predictions
for both models compared to the measured 3D power spectra.
The 3D power spectra have been divided by the linear power spectrum at the given
redshift to highlight the deviations from linear theory on small scales (high-$k$). 
In the bottom panel, we plot the residuals from the best-fit Arinyo model.
Cosmic variance causes large fluctuations on large scales (low-k), but the
residuals on intermediate and small scales ($k < 10 \ihMpc$) are within 5\%.



The ratio of both best-fit models is shown with dashed lines in the 
bottom panel of \cref{fig:ari_vs_mcd}.
The best-fit linear bias parameter from the Arinyo model ($b_F=-0.234$) is
3\% larger than the value from the McDonald model ($b_F=-0.227$).
As discussed in \cite{2015JCAP...12..017A}, the modelling of non-linear growth
in the McDonald model (an exponential term) extends to much larger scales
than the modelling in the Arinyo model.
The best-fit values for the RSD parameters are similarly consistent:
$\beta_F=1.34$ (Arinyo) and $\beta_F=1.31$ (McDonald).
The minimum values of pseudo-$\chi^2$ are also very similar, $\chi^2=48.1$ (Arinyo)
and $\chi^2=48.0$ (McDonald) with 225 data points.
\footnote{The pseudo-$\chi^2$ is defined in \cref{eq:pseudo_chi2}
and cannot be used as an indication of goodness of fit.}
Given that the Arinyo model has 3 fewer free parameters, we decide to use
that model in the following sections.

\subsection{Non-linear growth in the flux power spectrum}

\begin{figure}
    \centering
    \includegraphics[width=0.9\textwidth]{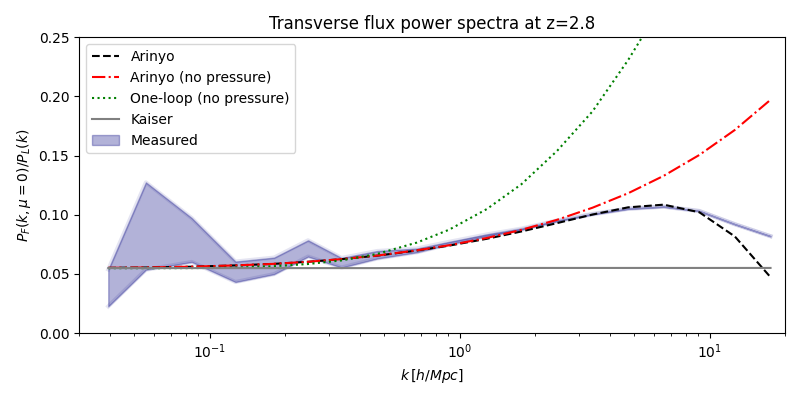}
    \caption{Comparison of different implementations
    of non-linear growth in the flux power spectrum,
    for the transverse modes ($\mu=0$).
    In this plot, only the standard Arinyo model (black dashed line) includes
    the effect of gas pressure, suppressing the power on scales of $k = 10 \ihMpc$.}
    \label{fig:p3d_lowk}
\end{figure}

Recently, \cite{2021JCAP...03..049G} presented an analysis of the 1D
correlations in the \lyaf\ using a model inspired by perturbation
theory and effective field theory of large-scale structure.
Instead of relying on free parameters to describe the non-linear growth of
structure on mildly non-linear scales, the authors used a model (first 
presented in \cite{2018JCAP...09..011G}) that captures non-linearities by 
using matter and velocity power spectra computed with perturbation theory at one-loop:
\begin{equation} \label{eq:one_loop}
 P_{\rm 3D}(k,\mu) = b^2 \left[ P^{\rm 1-loop}_{\delta\delta}(k) 
    + 2 \beta \mu^2 P^{\rm 1-loop}_{\delta \theta}(k) 
    + \beta^2 \mu^4 P^{\rm 1-loop}_{\theta \theta}(k) \right] D_{\rm IGM}(k,\mu) ~,
\end{equation}
where $P^{\rm 1-loop}_{\delta\delta}$, $P^{\rm 1-loop}_{\delta\theta}$, and
$P^{\rm 1-loop}_{\theta\theta}$ are the matter-matter, matter-velocity divergence, 
and velocity divergence-velocity divergence power spectra computed up to
1-loop in perturbation theory (they are equal at tree level).
$D_{\rm IGM}$ is a small-scale correction that models gas pressure with an
isotropic Gaussian suppression and thermal broadening with another Gaussian
suppression along the line of sight.
Even though the authors of \cite{2021JCAP...03..049G} only proposed this model
for analyses of the 1D power spectrum, here we discuss the interesting idea
of using perturbation theory at one-loop to model the mildly non-linear regime
of the 3D flux power spectrum.

Transverse modes ($\mu=0$) of the flux power spectrum are not affected by
redshift-space distortions and offer a cleaner path to study the effect
of non-linear growth of structure on the \lyaf.
As can be seen in \cref{fig:p3d_lowk}, the transverse flux power
spectrum at $z=2.8$ (blue band) rises above linear theory around $k=0.3\ihMpc$
until the effect of gas pressure suppresses the power around $k=7\ihMpc$.
The dashed black line shows the best-fit Arinyo model (again with $q_2=0$)
from \cref{fig:ari_vs_mcd};
the dot-dashed red line shows the same model without the pressure cut-off.
The solid gray line further sets $q_1=0$, effectively becoming a Kaiser model.
Last, the dotted green line shows the prediction from the one-loop model of
\cref{eq:one_loop} without the IGM correction when using the same
large scale bias parameters ($b_F$,$\beta_F$) as the other models.

As we will see in the next sections, these one-loop models (proposed initially by
\cite{2004PhRvD..70h3007S} in the context of galaxy clustering) are quite successful
in modeling the 3D power spectrum of halos.
The \lyaf, however, is less sensitive to the high-density regions of the box 
where the flux quickly saturates to $F=0$ and non-linearities in the flux power
spectrum grow slower than those in the matter power spectrum. 
As shown in \cref{fig:p3d_lowk}, this causes the one-loop model to over-predict
the growth of structure in the \lyaf.

In~\cite{2021JCAP...03..049G}, the impact of non-linearities on the 1D flux power spectrum is captured by adding a counterterm that parameterizes the impact of strong small-scale non-linearities, along the lines of effective field theory (EFT) methods.
For the 3D spectrum, the counterterm from~\cite{2021JCAP...03..049G} is not sufficient, since it describes the influence of non-linear corrections on the integrated 1D spectrum only, by construction. 
Instead, a more detailed EFT model would be required for the 3D spectrum, along the lines of \cite{Desjacques:2018pfv}.

Nevertheless, the one-loop model correctly describes the onset of deviations from the linear Kaiser formula, within the regime $k\lesssim 0.5 \ihMpc$ where these corrections are small.
This corresponds to the expected range of validity of perturbative methods. Increasing the range of agreement would require to include two-loop terms~\cite{2018JCAP...09..011G} along with corresponding counterterms~\cite{Baldauf:2015aha}. 
While being limited in $k$ range to weakly non-linear scales, the perturbative treatment has the advantage of being applicable also to non-standard cosmological models and can be evaluated efficiently for a large set of parameters.
It is possible to further improve the perturbative description in the BAO range by taking infrared resummation into account~\cite{Eisenstein:2006nj,Seo:2007ns,Baldauf:2015xfa}. 
It captures the impact of large-scale flows on the BAO feature in a systematic manner in real and redshift space~\cite{Blas:2016sfa,Ivanov:2018gjr}, leading to a broadening of the BAO peak in the correlation function, and a damping of the BAO oscillations in the power spectrum. 
The IR resummation can easily be implemented in the perturbative model by replacing the linear spectrum that enters the loop computation by a modified input spectrum, obtained by decomposing $P^\text{lin}=P^\text{nw}+P^\text{w}$ into a smooth (non-wiggly) and an oscillating (wiggly) contribution, and multiplying the latter with a particular Gaussian damping factor~\cite{Blas:2016sfa,Ivanov:2018gjr}. 
While IR resummation is important within the BAO range, it does not change the behaviour of perturbative predictions of the power spectrum in a sizeable way at large wavenumbers $k\gtrsim 0.5 \ihMpc$ shown in \cref{fig:p3d_lowk}.

\section{Modelling the halo power spectrum}
\label{sec:halo}

\subsection{Real-space halo power spectrum} \label{ss:halo_no_RSD}

\begin{figure}
    \centering
    \includegraphics[width=0.9\textwidth]{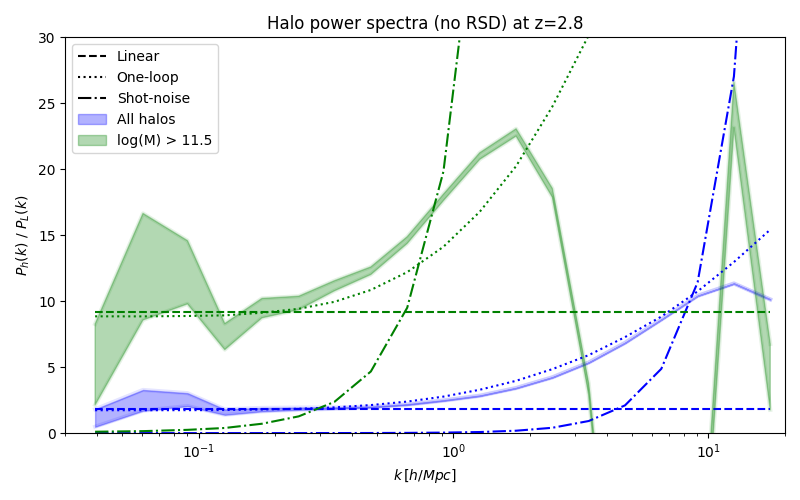}
    \caption{Real space (no RSD) halo power spectrum measured at $z=2.8$ from the whole halo
    catalog (blue) and from the high-mass bin (green), divided by the linear power spectrum.
    Dashed (dotted) lines show the best-fit linear (one-loop) model,
    fitted to $k_{\rm max}= 0.3 \ihMpc$.
    The shot-noise has been subtracted from both the measurement and the models, and it is
    shown in dot-dashed lines.}
    \label{fig:halo_no_rsd}
\end{figure}

The real-space halo power spectrum as a function of halo mass is shown in
the right panel of \cref{fig:halo_masses}, together with its best-fit linear model
and the shot-noise contribution.
In \cref{fig:halo_no_rsd}, we show this again for the largest mass bin (green) and
for the total halo catalog (blue), but this time divided by the linear power spectrum of
matter fluctuations. 
Dashed lines show the best-fit linear model, using only large scales ($k_{\rm max}=0.3 \ihMpc$).
Dotted lines show the best-fit model when we use a one-loop matter density power spectrum instead. 
All models and measurements are shown after subtracting the shot-noise, shown
as dot-dashed lines for each sample. 

Here we model the shot-noise power as the inverse of the halo density.
This simple approximation is clearly over-subtracting power in the case of the high-mass
halo sample, since the shot-noise subtracted power spectrum becomes negative at $k > 3 \ihMpc$.
A detailed study of the role of shot-noise in halo clustering can be found in \cite{2017PhRvD..96h3528G}.

\subsection{Redshift-space halo power spectrum}

\begin{figure}
    \centering
    \includegraphics[width=0.9\textwidth]{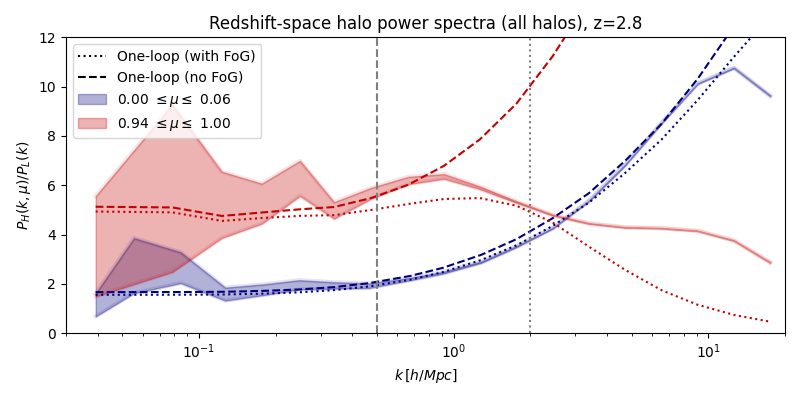}
    \caption{Comparison of the redshift-space halo power spectrum measured at $z=2.8$ along (red)
    and transverse to (blue) the line of sight.
    Dotted lines show the best-fit one-loop model (\cref{eq:sco}),
    including a Lorentzian term describing fingers of God (\cref{eq:D_FoG}),
    when fitted to $k_{\rm max}=2 \ihMpc$.
    Dashed lines show the best-fit one-loop model when using only $k_{\rm max}=0.5 \ihMpc$
    and ignoring FoG.
    The shot-noise has been subtracted from both the measurement and the models.}
    \label{fig:sco_lowk}
\end{figure}

The ubiquitous model for the clustering of galaxies (or halos) in redshift space
was presented by Kaiser \cite{1987MNRAS.227....1K}:
\begin{equation}\label{eq:kaiser_halo}
    P_\textrm{h}(k,\mu)=b_h^2(1+\beta_h\mu^2)^2P_L(k).
\end{equation}
In this equation, $b_h$ is the linear halo bias and $\beta_h \equiv f/b_h$ is the linear RSD parameter for halos. This linear theory description works well on large scales but becomes less accurate past $k \gtrsim 0.5 \ihMpc$ at $z=2.8$. 
An alternative description of galaxy clustering using non-linear power spectra
(equivalent to \cref{eq:one_loop} without $D_{\rm IGM}$) was presented in \cite{2004PhRvD..70h3007S}:
\begin{equation}\label{eq:sco}
    P_\textrm{h}(k,\mu)=b_h^2\Big[P^{\rm 1-loop}_{\delta\delta}(k) 
        + 2f\mu^2 P^{\rm 1-loop}_{\delta\theta}(k) 
        + f^2\mu^4 P^{\rm 1-loop}_{\theta\theta}(k)\Big] ~.
\end{equation}

Non-linear peculiar velocities of halos, often referred to as ``Fingers of God'' (FoG),
suppress the halo power spectrum along the line-of-sight direction.
This is often modeled using a Gaussian or Lorentzian distribution as a multiplicative factor to the above equations.\footnote{See \cite{2004PhRvD..70h3007S} for a detailed explanation addressing why this simplified model cannot be correct.} 
In particular, we use a correction of the form
\begin{equation}\label{eq:D_FoG}
    D_\textrm{FoG}(k,\mu) \equiv \frac{1}{1+(k\mu\sigma_v)^2},
\end{equation}
where $\sigma_v$ is an effective pairwise velocity dispersion parameter \cite{2009MNRAS.393..297P}. 

In \cref{fig:sco_lowk} we plot the line-of-sight (red) and transverse (blue) halo
power spectrum measured at $z=2.8$, when using all halos in the catalog, and divided 
by the linear power at that redshift.
Dashed lines show the best-fit model when using only the largest scales
($k_{\rm max}=0.5 \ihMpc$) and ignoring FoG, while the dotted lines show the best-fit
model including Lorentzian FoG and fitting scales to $k_{\rm max}=2 \ihMpc$. 
The transverse modes are remarkably well fitted to high $k$, 
but the line-of sight modes are not well modelled with this simple FoG term.
For this reason, in the rest of this paper we do not attempt to model the redshift-space
halo power spectrum, and we obtain the large-scale halo bias from the real-space halo
power spectrum (\cref{fig:halo_no_rsd}).

\section{Modelling the \lya\ forest-halo cross-power spectrum}
\label{sec:cross}

\begin{figure}
    \centering
    \includegraphics[width=0.9\textwidth]{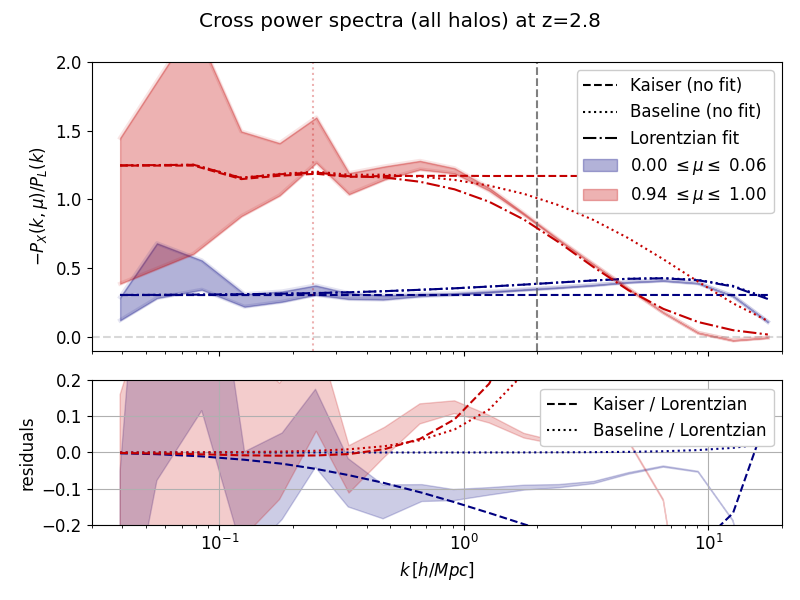}
    \caption{(Top) Measured \lyax\ cross-power spectra when using all halos in the snapshot (shaded regions),
    divided by the linear power spectrum, as a function of wavenumber $k$ and
    orientation with respect to the line of sight $\mu$.
    Dashed and dotted lines show the predictions based on the Kaiser model and on
    \cref{eq:baseline_cross} respectively, both without any free parameter.
    Dot-dashed lines show the best-fit model after adding an
    extra parameter $\sigma_v$ to account for fingers of God with a Lorentzian term, using wavenumbers smaller than $k_{\rm max}=2 \ihMpc$,
    marked with a vertical dashed gray line.
    Vertical color dotted lines show the scales where quasar redshift
    errors of $500 ~ \kms$ would suppress half of the power.
    (Bottom) Residuals with respect to the best-fit Lorentzian model.}
    \label{fig:cross_all}
\end{figure}

In \cref{fig:cross_all} we show the same \lyax\ power spectra that is also shown
in \cref{fig:measured_p3d}, but this time divided by the linear power spectrum of 
matter fluctuations at the same redshift. 
In this section we discuss models to describe the non-linearities in this measurement.

\subsection{Linear model}

Given that the Kaiser model can describe the large-scale clustering of both tracers (halos and \lyaf), 
and that these are strongly anti-correlated on large scales (see bottom panel of
\cref{fig:measured_p3d}), one expects that the Kaiser model should also describe the 
\lyax\ cross-correlation on linear scales.
This model can be computed as the geometric mean of the Kaiser models for both tracers, 
\cref{eq:kaiser_flux,eq:kaiser_halo}:
\begin{equation}\label{eq:linear_cross}
  P_\textrm{Fh}(k,\mu) = \sqrt{ P_\textrm{F}(k,\mu) ~ P_\textrm{h}(k,\mu) }
    = b_h(1+\beta_h\mu^2) ~ b_F(1+\beta_F\mu^2) ~ P_L(k) ~.
\end{equation}
Indeed, \cref{fig:cross_all} shows that the large-scale cross-power is proportional to
the linear power spectrum, with a boost in line-of-sight modes caused by linear RSDs.

This linear theory model was used in the first measurement of the \lya-quasar cross correlation
in BOSS at $z=2.3$ over scales $r>15 \hMpc$ \cite{2013JCAP...05..018F}.
The authors also included a multiplicative correction, similar to \cref{eq:D_FoG}, to model the
suppression of power caused by quasar redshift errors and non-linear quasar velocities.
They obtained values for the velocity dispersion of the order of $500\, \kms$, and argued
that these were dominated by quasar redshift errors.

\subsection{Building nonlinear models}

BAO measurements using the cross-correlation of quasars and the \lyaf\ from BOSS and eBOSS
\cite{2014JCAP...05..027F,2017A&A...608A.130D,2019A&A...629A..85D,2019A&A...629A..86B,2020ApJ...901..153D}
used a modified version of \cref{eq:linear_cross}.
They replaced $P_L(k)$ by a quasi-linear power spectrum that decoupled the BAO peak from the
smooth component in order to model the non-linear broadening of the BAO peak \cite{2013JCAP...03..024K}
(see also discussion at the end of \cref{sec:flux}).
However, given the limited box size of our simulations, we do not attempt to model this
non-linear broadening of the BAO wiggles, and we will focus here on the deviations from
linear theory on small scales. 

In order to build our first non-linear model for the \lyax\ correlations, 
we use again the geometrical mean of the two power spectra. 
This time, however, we use the Arinyo model to describe the flux component:
\begin{equation}\label{eq:baseline_cross}
  P_\textrm{Fh}(k,\mu) = b_h(1+\beta_h\mu^2) ~ b_F(1+\beta_F\mu^2) ~ P_L(k) 
        ~ \sqrt{D_{\rm NL}(k,\mu)} ~,
\end{equation}
where $D_{\rm NL}(k,\mu)$ is the small-scales correction described in \cref{eq:D_NL}.
We will refer to this equation as the \textit{baseline} model for the \lyax\ cross-correlation.

Instead of fitting this model to the measured cross-correlation, we start by making predictions
when using the best-fit values measured from the \lyaf\ auto-correlation in \cref{sec:flux}, 
and the halo bias measured from the halo auto-correlation in \cref{sec:halo}.
Because of the difficulties in modelling RSDs in the halo power spectrum, 
we use instead the halo bias measured from the real space halo power spectrum
(see \cref{fig:halo_no_rsd}).
The dotted lines in \cref{fig:cross_all} show the predicted cross-correlation using 
\cref{eq:baseline_cross}.
Unfortunately, this simple model (without any free parameter) does not seem to perform any
better than the Kaiser model, and it overpredicts the signal in the transverse direction (blue lines).

In order to improve the model for the line-of-sight signal (in red), we also add a fit
that includes an extra term to model Fingers-of-God (FoG) in the halos:
\begin{equation}\label{eq:cross_FoG}
  P_\textrm{Fh}(k,\mu) = b_h(1+\beta_h\mu^2) ~ b_F(1+\beta_F\mu^2) ~ P_L(k) 
        ~ \sqrt{D_{\rm NL}(k,\mu) ~ D_{\rm FoG}(k,\mu) } ~,
\end{equation}
where $D_{\rm FoG}(k,\mu)$ is the Lorentzian function from \cref{eq:D_FoG}.
The dot-dashed lines in the top panel of \cref{fig:cross_all} show the best-fit model when
fitting $\sigma_v$ using wavenumbers smaller than $k_{\rm max} = 2 \ihMpc$
(gray dashed vertical line).

The bottom panel of \cref{fig:cross_all} shows the residuals for the
\lyax\ power when using the Baseline model (no free parameters), as well as
the ratio of the different models.
The simple Kaiser model of \cref{eq:linear_cross} properly describes the
measured correlations up to $k=1 \ihMpc$, with an accuracy better than 5\%.
The other models, however, seem to overpredict the clustering of the
transverse modes (blue lines).

It is important to keep in mind that measurements of the cross-correlation
of quasars and the \lyaf\ are affected by quasar redshift errors.
Assuming a Gaussian suppression with $\sigma_v = 500 ~ \kms$, typical of
quasar surveys at $z>2$ \cite{2020ApJ...901..153D}, we compute
the wavenumber where half of the power would be suppressed. 
These are shown as vertical colored dotted lines, as a function of $\mu$.
Disagreements on scales much smaller than these are not necessarily relevant,
since they will be highly suppressed by quasar redshift errors when analysing real data.

\subsection{Dependency on halo mass} \label{sec:cross_mass}

\begin{figure}
    \centering
    \includegraphics[width=0.9\textwidth]{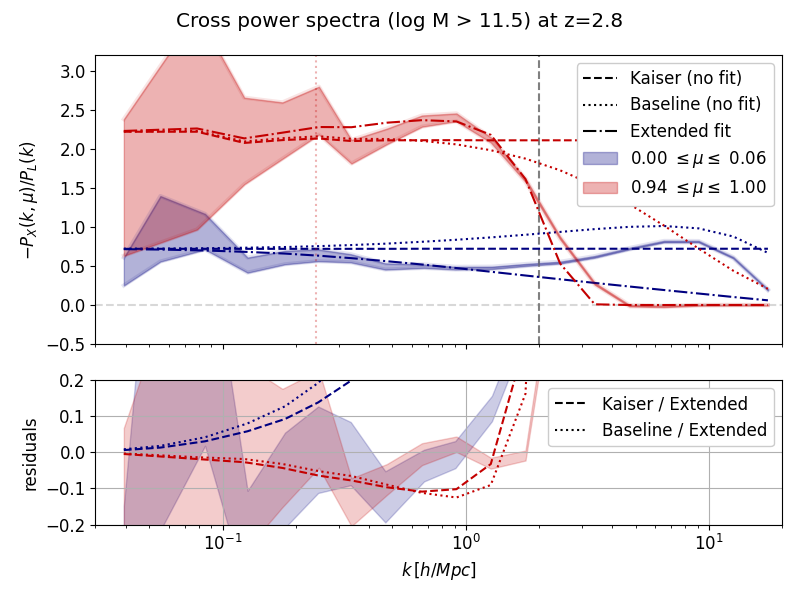}
    \caption{Same as \cref{fig:cross_all}, but limited to the most massive halos.
    Dashed and dotted lines show the predictions based on the Kaiser model and on
    \cref{eq:baseline_cross} respectively, both without any free parameter.
    Dot-dashed lines show the best-fit model after adding the three extra parameters ($\alpha$, $\gamma$, $\nu$) in our extended model.
    The fit uses wavenumbers smaller than $k_{\rm max}=2 \ihMpc$,
    marked with a vertical dashed gray line.}
    \label{fig:cross_highM}
\end{figure}

As discussed at the end of \cref{sec:halo_grids}, most of the halos in our simulation
are significantly smaller than those expected to host quasars.
Therefore, the measurements of the cross-correlation presented above might
not be representative of the cross-correlation with quasars. 
However, the green lines in \cref{fig:halo_no_rsd} show that the clustering
of the most massive halos in the simulation is strongly affected by shot-noise
on scales of $k > 0.5 \ihMpc$.

In \cref{fig:cross_highM} we show the \lyax\ cross-power when looking
at halos in the largest mass bin.
We can see that the baseline prediction of \cref{eq:baseline_cross} (dotted lines)
overpredicts the clustering already on fairly linear scales ($k < 0.3 \ihMpc$).
These deviations are also present at $\mu=0$ (blue lines), and cannot be related
to non-linear RSDs or FoG.
If one used the one-loop power spectrum instead of the linear one 
in \cref{eq:baseline_cross}, the deviations would be even larger.

As can be seen in the left panel of \cref{fig:cross_issues}, massive halos are less
(anti) correlated with the \lyaf, particularly for transverse modes where the correlation
coefficient is only $\rho=-0.7$ at $k=0.3 \ihMpc$, and $\rho=-0.5$ at $k=0.5 \ihMpc$.
Therefore, instead of trying to improve the non-linear modelling, we developed a more 
complex (\textit{Extended}) model that adds to \cref{eq:baseline_cross} an ad-hoc
multiplicative correction to capture the anisotropic decorrelation of both fields:
\begin{equation}\label{eq:D_M}
    D_\textrm{M}(k,\mu)  \equiv \exp\Big[(\alpha+\gamma\mu^2) \Delta^2(k) -(k\mu\nu)^4\Big],
\end{equation}
where $\alpha$, $\gamma$, and $\nu$ are free parameters, and $\Delta^2(k)$ 
is the dimensionless linear power as described in \cref{eq:delta2}.
The dot-dashed lines in \cref{fig:cross_highM} show the best-fit predictions for 
this extended model, 
and the bottom panel shows the residuals with respect to this fit.
In this case the residuals are smaller than 10\% up to scales of $k=1 \ihMpc$.

It is important to remember, however, that the measurement on these scales
is probably impacted by the \texttt{Quick-Lya} approximation used in the simulations.
As shown in \cite{2017MNRAS.468.1893M}, the impact of the approximation is more important
for massive halos with large virial radius.




\section{Discussion}
\label{sec:discussion}

\subsection{Prediction of 1D correlations from 3D models}

In \cref{sec:flux} we have shown that the \textit{Arinyo} model
can successfully describe the full shape of 3D correlations down
to small scales. 
Integrating the 3D model with \cref{eq:p1d}, one can also obtain
a prediction for the 1D correlations:
\begin{equation}
 \label{eq:p1d}
 P_{\rm 1D}(k_\parallel) = \frac{1}{\pi} \int_0^\infty d k_\perp 
            ~ k_\perp ~ P_{\rm 3D}(k_\parallel,k_\perp) ~,
\end{equation}
where $k_\parallel = k \mu$, and $k_\perp = k \sqrt{1 - \mu^2}$.
The 1D power spectrum measured at a given line-of-sight wavenumber $k_\parallel$ is
therefore sensitive to the 3D power spectrum over all transverse modes $k_\perp$.

\begin{figure}
    \centering
    \includegraphics[width=0.9\textwidth]{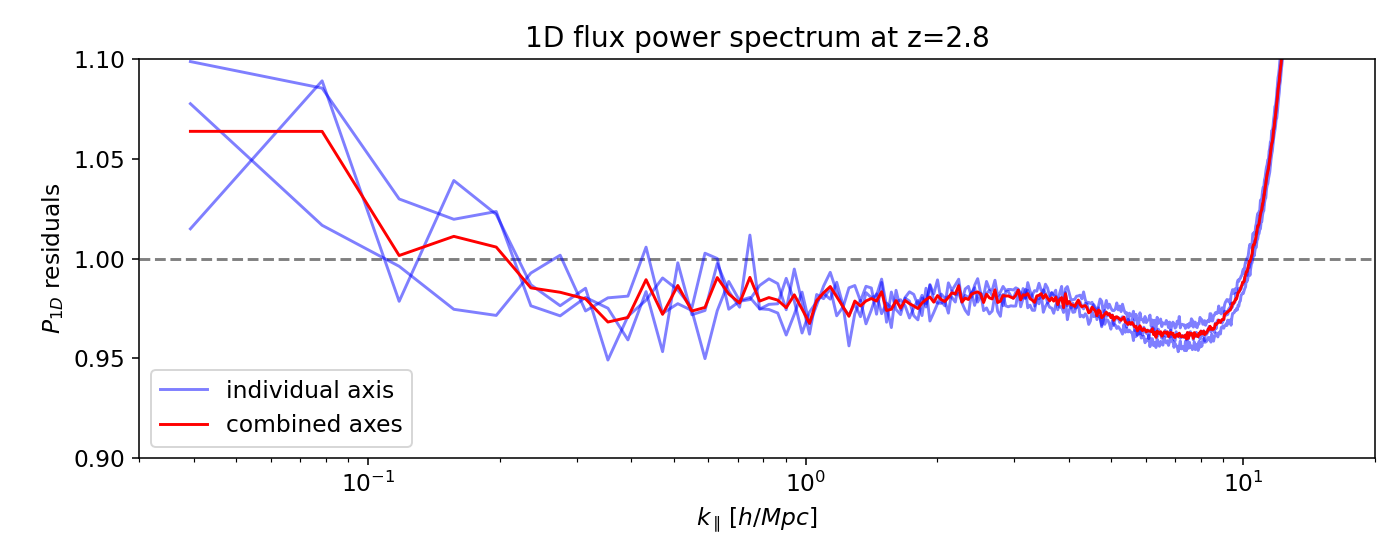}
    \caption{One-dimensional flux power spectrum measured along each
    simulation axis (blue), and its mean (red), divided by the
    prediction from the best-fit Arinyo model from \cref{sec:flux}.}
    \label{fig:p1d_residuals}
\end{figure}

The blue lines in \cref{fig:p1d_residuals} show the P1D measurement from
\lyaf\ skewers along the three different simulation axes, divided by
the prediction from the best-fit Arinyo model in \cref{sec:flux}.
The red line is the mean of the three blue lines, and is our best
measurement of 1D correlations from the simulation.
The $6\%$ differences on the largest scales might be explained by 
cosmic variance, and the deviations on smaller scales are less than $2\%$ on
scales relevant for BOSS, eBOSS or DESI analyses ($k_\parallel < 4 \ihMpc$).

\subsection{Impact on BAO studies}

\begin{figure}
    \centering
    \includegraphics[width=0.48\textwidth]{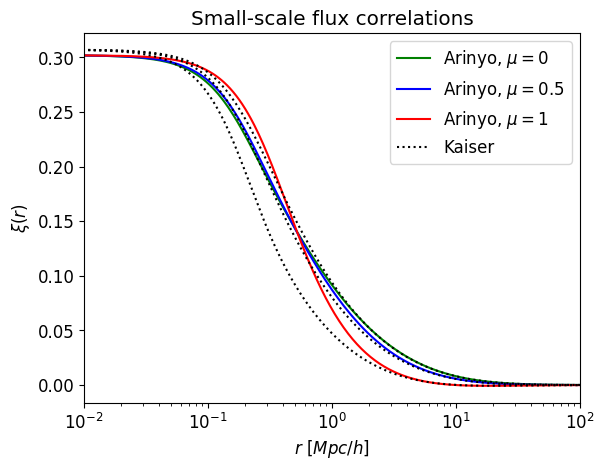}
    \includegraphics[width=0.48\textwidth]{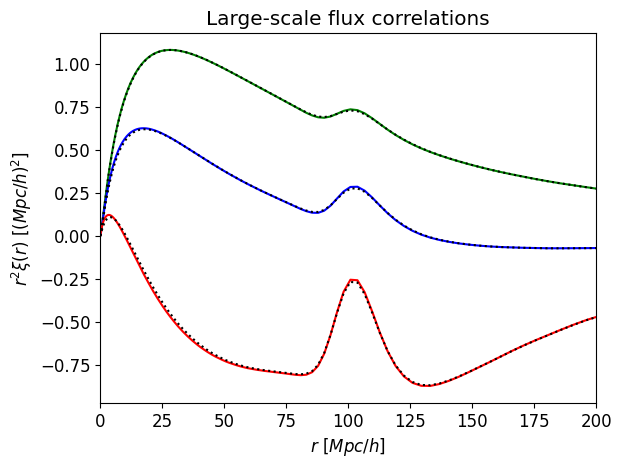}
    \includegraphics[width=0.48\textwidth]{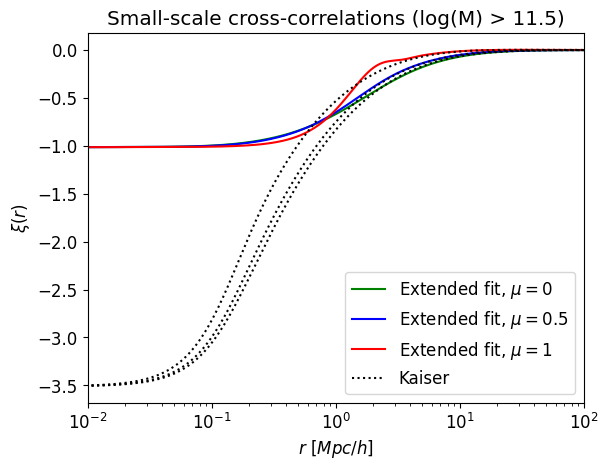}
    \includegraphics[width=0.48\textwidth]{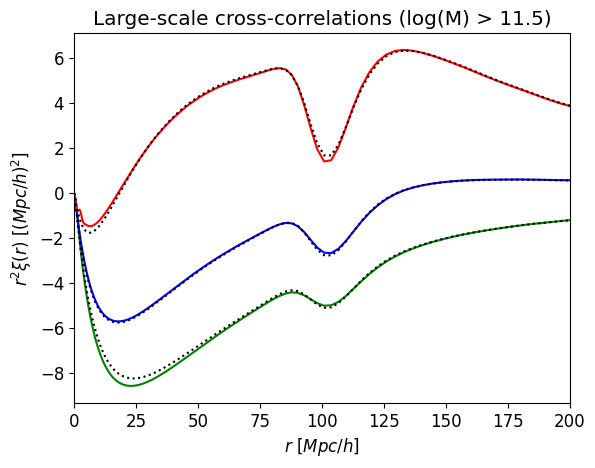}    
    \caption{Correlation functions corresponding to the best-fit models obtained in
    previous sections, for different orientations with respect to the line of sight $\mu$.
    Solid color lines show the best-fit Arinyo model for the \lyaf\ auto-correlation 
    shown in \cref{fig:ari_vs_mcd} (top panels); 
    the best-fit extended model for the \lyax\ cross-correlation when using only halos
    in the largest mass bin, as in \cref{fig:cross_highM} (bottom panels).
    Black dotted lines show the equivalent correlations for the corresponding
    \textit{Kaiser} model, using the same values of the linear bias parameters.
    Left panels show correlations on small separations, while the right panels
    show large-scale correlations multiplied by $r^2$ to highlight the BAO feature. }
    \label{fig:xi_bao}
\end{figure}

So far, we have presented measurements and models of power spectra, 
correlations in Fourier space.
However, BAO studies using the \lyaf\ to date have used the correlation function.
In \cref{fig:xi_bao} we show the model correlations in configuration space
for the flux auto-correlation (top panels), and for the \lyax\ cross-correlation
when using the most massive halos (bottom panels).
Solid color lines show the correlations for the main \textit{Arinyo} model (top panels),
for different orientations with respect to the line of sight $\mu$, and for the
extended cross-correlation model from \cref{sec:cross_mass} (bottom panels).
Black dotted lines show the equivalent correlations for the corresponding 
\textit{Kaiser} model, where we have set $D_{NL}=1$ and used the same
values of the linear bias parameters.

In order to compute these model correlations, we have computed the predicted
multipoles of the power spectra up to the 10th order in the Legrendre
decomposition, and we have computed their inverse Fourier transform using
the FFTLog formalism as implemented in \texttt{mcfit} 
\footnote{ \url{https://github.com/eelregit/mcfit}. } .

The differences between the models are clear when looking on small scales
(left panels), but on the large scales used in BAO analyses (right panels) the
differences are significantly smaller, especially for the flux auto-correlation.
Recently, \cite{2021MNRAS.506.5439C} has proposed using the anisotropy of these
correlations to constraint cosmology by performing an Alcock-Paczi\'nsky test.
\Cref{fig:xi_bao} shows that deviations from linear theory should be relatively
small if one considers only separations larger than $30 \hMpc$ or so.

\subsection{Problems modelling the cross-correlation for massive halos}

As discussed in \cref{sec:halo_grids}, the limited box size of the simulations
results in relatively few halos with masses comparable to those hosting quasars.
More importantly, as discussed in \cref{sec:sim_details}, the \texttt{Quick-Lya}
approximation used in this and many other simulations of the \lyaf\ artificially
removes gas from the high-density regions and biases the measurements of the
\lyax\ correlation on small scales. 
Besides these issues with the measurements, in this section we discuss three other
problems that we encounter when modelling the \lyax\ cross-correlation. 

First, as discussed in appendix B of \cite{2012JCAP...11..059F}, 
the cross-correlation of halos with the \lyaf\ can be computed as 
the mean value of the \lya\ fluctuations at a given separation from a halo:
\begin{equation}
 \xi_X(r) = \langle \delta_F \rangle_r   ~,
\end{equation}
where the expected value is computed using only pixels at a separation $r$.
In the absence of instrumental noise, the value of flux can not be negative.
This implies that the cross-correlation always has to satisfy $\xi_X(r) \geq -1$. 
However, our power spectrum models do not know about this natural threshold.
As can be seen in the bottom-left panel of \cref{fig:xi_bao},
it seems that our extended model asymptotically reaches $\xi_X=-1$.
However, we have tested that small variations in the model result in
non-physical correlations with $\xi_X(r) < -1$.

\begin{figure}
    \centering
    \includegraphics[width=0.45\textwidth]{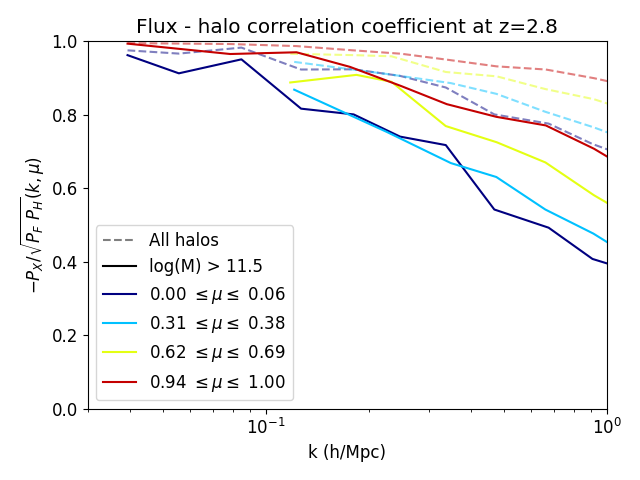}
    \includegraphics[width=0.45\textwidth]{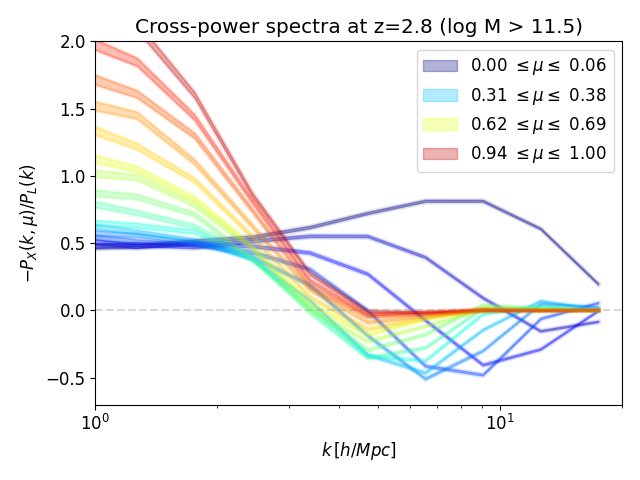}    
    \caption{Left: Correlation coefficient between halos and the \lyaf, 
    for the most massive halos (solid lines) and for the complete catalog (dashed).
    Right: Cross-correlation of the most massive halos on the smallest scales,
    divided by the linear power spectrum.}
    \label{fig:cross_issues}
\end{figure}

Second, the models discussed in \cref{sec:cross} are motivated by the assumption
that the \lyaf\ field and the halo density are strongly (anti-) correlated, and
try to build a model from the geometrical mean of auto-correlation models.
However, in the left panel of \cref{fig:cross_issues} we compare the correlation
coefficient when using all halos (dashed lines, also seen in \cref{fig:measured_p3d})
and when using only the most massive halos (solid lines).
For the massive halos, half of the anti-correlation has already been lost on
scales of $k=0.5 \ihMpc$.
The ad-hoc correction from \cref{eq:D_M} can capture this by fitting negative 
values for $\alpha$, but it is difficult to propose physically-motivated models
when the fields become increasingly uncorrelated.

Finally, our models assume that the cross-power spectrum is negative on all scales.
However, the right panel of \cref{fig:cross_issues} clearly shows that the 
cross-power for the most massive halos changes sign around $k = 3 \ihMpc$, 
except for the most transverse modes (in dark blue).
This might be a consequence of the \texttt{Quick-Lya} approximation used in the
simulations, and it would be interesting to carry out an equivalent study of these
small-scale correlations in configuration space.

\section{Conclusions}
\label{sec:conclusions}

We have presented the most accurate measurement of the 3D power spectrum of 
fluctuations in the \lyaf\ to date, using outputs from the Sherwood suite
of hydrodynamic simulations 
\footnote{ \url{https://www.nottingham.ac.uk/astronomy/sherwood} }.
These simulations are considerably larger (20 times more particles) than those
used in the current state-of-the-art measurements of \cite{2015JCAP...12..017A}.
We have also presented, for the first time, a measurement of the \lyax\ 3D cross-power spectrum.
The \lyaf\ and the halo catalog are strongly anti-correlated on all scales
relevant for BAO analyses ($\rho < -0.9$ at $k < 0.3 \ihMpc$).
The most massive halos, however, are less anti-correlated (left panel of \cref{fig:cross_issues}).
We have made these measurements public 
\footnote{ \url{https://github.com/andreufont/sherwood_p3d} }
to encourage other teams to work on the modelling of these correlations.
  
We have confirmed that the non-linear models of \cite{2003ApJ...585...34M} and 
\cite{2015JCAP...12..017A} are able to describe the 3D flux power spectrum down to
small scales, and that the second-order non-linear growth parameter ($q_2$) in
\cite{2015JCAP...12..017A} is not required to obtain a good fit.
In \cref{fig:ari_vs_mcd,fig:p1d_residuals} we show that the Arinyo model can predict at the same
time 1D and 3D correlations, with 3D residuals smaller than 5\% on scales smaller
than $k < 10 \ihMpc$, and 1D residuals smaller than 2\% for $k < 4 \ihMpc$.
This opens the possibility of joint fits of 1D and 3D correlations observed by BOSS and
eBOSS, and the possibility of building 1D emulators based on 3D models.
The non-linearities in the \lyaf\ power spectrum are smaller than those in the
matter power spectrum (see \cref{fig:p3d_lowk}).
Dense regions in the simulation quickly saturate the transmitted flux fraction to $F=0$,
acting as a natural shield that protects the \lyaf\ from the details of non-linear growth.
  
The bottom panel of \cref{fig:cross_all} shows that the Kaiser model
describes the \lyax\ power spectrum with residuals smaller than 5\% even
on scales as small as $k=1 \ihMpc$, when using all halos in the simulation.
However, it does not describe well the correlation with the most massive 
halos, representative of quasars.
In \cref{eq:D_M} we present an ad-hoc multiplicative correction to the model
that can improve the fit considerably, with residuals smaller than 10\% up to $k < 1 \ihMpc$.

Even though the Sherwood suite is the state of the art in terms of
simulations of the \lyaf, in \cref{app:alt} we show that the measurements
presented here are still affected by their limited box size and resolution.
Uncertainties in the cosmological model assumed in the simulation, as well
as in the assumed thermal and ionization history, could also modify 
significantly the measured power spectra.
For this reason we do not focus on the actual best-fit parameter values that
we obtain when analysing this particular simulation;
as we show in \cref{app:alt}, both the large scale bias and the pressure terms
are affected by the limitations of the simulations.

In \cref{sec:discussion} we list a couple of fundamental issues with our
approach to model the \lyax\ cross-correlation.
By definition, the cross-correlation function cannot be more negative than $\xi=-1$
(corresponding to $F=0$ or $\delta_F=-1$), but our Fourier-space models are not
aware of this.
A second limitation of our models is that they are not allowed to change sign,
while the measured \lyax\ cross-power seems to change sign around $k=3 \ihMpc$
as shown in the right panel of \cref{fig:cross_issues}.  
We caution the reader that the measurement itself can be corrupted in these small scales.
The \texttt{Quick-Lya} approximation used in the simulations artificially removes gas
from the vicinity of the dark matter halos.

Finally, in \cref{fig:xi_bao} we show that on scales relevant for BAO analyses,
the best-fit non-linear models predict small deviations from linear theory.
This is particularly true for the \lyaf\ auto-correlation, 
while larger deviations in the cross-correlation with the most massive halos
appear on scales below $r=30 \hMpc$ for the most transverse separations.

This investigation is timely since the Dark Energy Spectroscopic Instrument 
(DESI \cite{2016arXiv161100036D}) started its five-year program in May 2021, 
and it will soon provide the largest \lyaf\ dataset to date.
While the main goal of the DESI \lya\ dataset is to measure BAO, this study is a step towards cosmological analyses using the full shape of the 3D
correlations.

\acknowledgments
We would like to thank James Bolton for useful discussions and assistance with the Sherwood simulations.
The Sherwood simulation suite was performed with supercomputer time awarded by the Partnership for Advanced Computing in Europe (PRACE) 8th call. The Sherwood team also acknowledges use of the DiRAC Data Analytic system at the University of Cambridge, operated by the University of Cambridge High Performance Computing Service on behalf of the STFC DiRAC HPC Facility (www.dirac.ac.uk). This equipment was funded by BIS National E-infrastructure capital grant (ST/K001590/1), STFC capital grants ST/H008861/1 and ST/H00887X/1, and STFC DiRAC Operations grant ST/K00333X/1.
Early stages of this analysis used computing equipment funded by the Research Capital Investment Fund (RCIF) provided by UKRI, and partially funded by the UCL Cosmoparticle Initiative.
The final computation of 3D grids and power spectrum measurements were done at the Port d'Informaci\'o Cient\'ifica (PIC), a scientific-technological center maintained through a collaboration of the Institut de F\'isica d'Altes Energies (IFAE) and the Centro de Investigaciones Energ\'eticas, Medioambientales y Tecnol\'ogicas (CIEMAT).

JG acknowledges support from the U.S. Department of Energy Office of Science Graduate Student Research Fellowship Program, the Cosmology and Astroparticle Student and Postdoc Exchange Network, and Princeton's Presidential Postdoctoral Research Fellowship.
AFR acknowledges support by an STFC Ernest Rutherford Fellowship, grant reference ST/N003853/1, and funds trough the program Ramon y Cajal (RYC-2018-025210) of the Spanish Ministry of Science and Innovation. IFAE is partially funded by the CERCA program of the Generalitat de Catalunya.
CP acknowledges support by NASA ROSES grant 12-EUCLID12-0004.
The Dunlap Institute is funded through an endowment established by the David Dunlap family and the University of Toronto. 
MG is supported by the Excellence Cluster ORIGINS, which is funded by the Deutsche Forschungsgemeinschaft (DFG, German Research Foundation) under Germany’s Excellence Strategy - EXC-2094 - 390783311.
DB is supported by a `Ayuda Beatriz Galindo Senior' from the Spanish `Ministerio de Universidades', grant BG20/00228. 
The research leading to these results has received funding from the Spanish Ministry of Science and Innovation (PID2020-115845GB-I00/AEI/10.13039/501100011033).
VI is supported by the Kavli Foundation.
This work was partially enabled by funding from the UCL Cosmoparticle Initiative.

\bibliographystyle{JHEP.bst}
\bibliography{refs}

\appendix
\section{Alternative analyses of flux power spectrum}
\label{app:alt}

\subsection{Convergence of simulation boxes}

\begin{figure}
    \centering
    \includegraphics[width=0.48\textwidth]{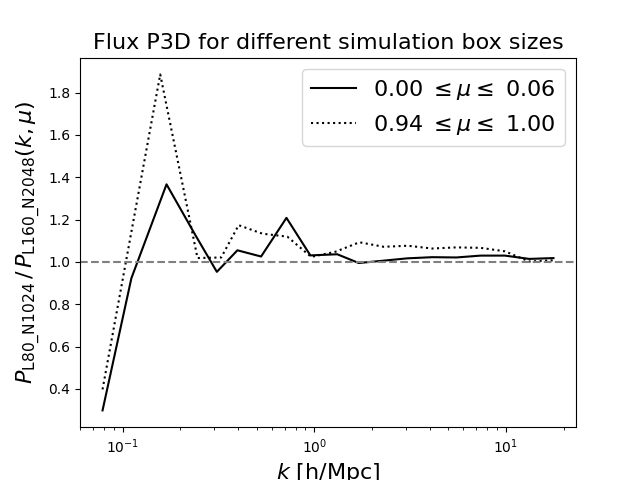}
    \includegraphics[width=0.48\textwidth]{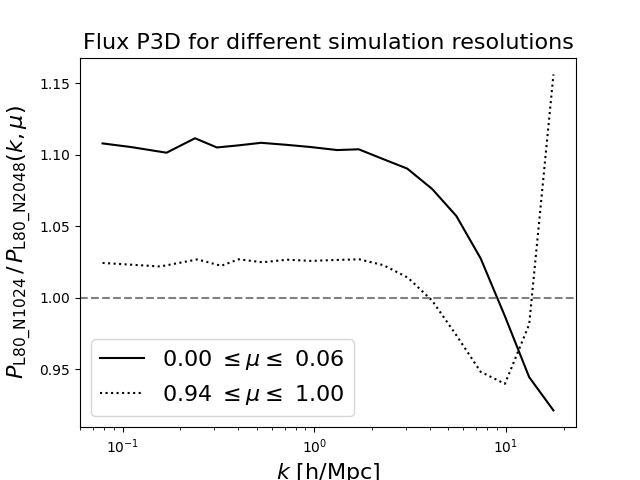}
    \caption{Left: Ratio of \lyaf\ power spectrum in two simulations with the same particle resolution, but different box size.
    The differences at low-k are probably explained by cosmic variance.
    The computation of the ratio on all scales is affected by interpolation artifacts, since the values of wavenumbers $k$ in the two measurements do not match.
    Right: Ratio of \lyaf\ power spectrum in two simulations with the same box size, but different particle resolution.} 
\label{fig:compare_box}
\end{figure}

Most measurements discussed in the main text use the largest simulation
\texttt{L160\_N2048}, with box size $L=160 \hMpc$ and $N=2048^3$ 
CDM and gas particles.
In order to study the impact of box size, in the left panel of
\cref{fig:compare_box} we compare the measurement from this box
with the measurement from a simulation that has the same particle
resolution, but a smaller box size (\texttt{L80\_N1024}).
Sample variance due to the limited box sizes makes it difficult to compare
the large-scale measurement, but one can see small differences extending
to all scales.
Note, however, that this ratio is affected by interpolation artifacts, 
since the values of wavenumbers $k$ in the two measurements do not match.

In the right panel of \cref{fig:compare_box} we study the impact of
particle resolution, using two simulations with $L=80 \hMpc$, but different
number of particles. 
The simulation \texttt{L80\_N1024} has the same particle resolution as
the box used in the main text, while the simulation \texttt{L80\_N2048}
has 8 times more particles in the same volume.
On large scales, particle resolution impacts the value of the large scale
bias and redshift space distortion parameter, at the order of a few percent.
On small scales, the gas in the simulation with lower particle resolution
seems to have a stronger pressure smoothing.

\subsection{Convergence of grids}

Memory restrictions in the post-processing of the simulations limit our
analysis to a 3D grid of $(1024)^2 \times 2048$ cells, with a cell
size of $0.156 \hMpc$ in the transverse direction and $0.078 \hMpc$ along the
line of sight.
Here we study the impact of this pixelisation in the flux power, 
using the \texttt{L80\_N1024} simulation, with a smaller box
but the same resolution \edits{as} our largest simulation.

\begin{figure}
    \centering
    \includegraphics[width=0.49\textwidth]{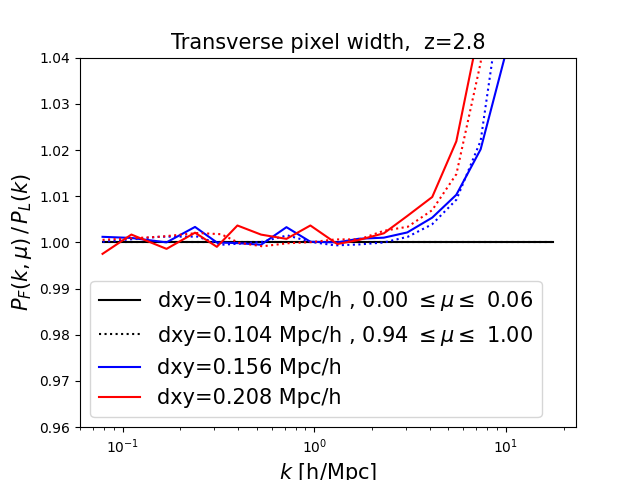}
    \includegraphics[width=0.49\textwidth]{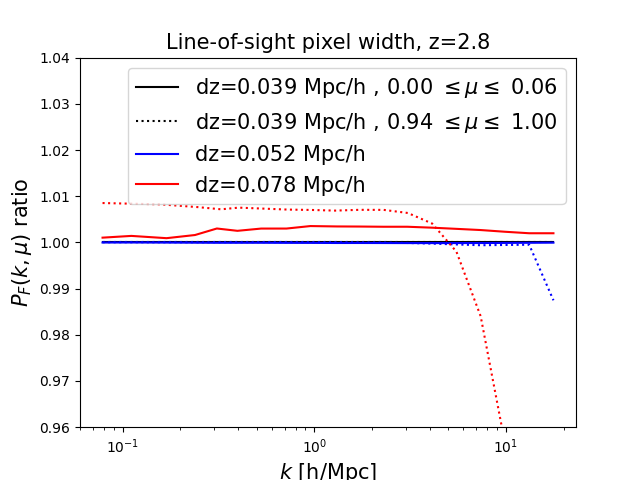}
    \caption{Measured flux power spectrum at $z=2.8$, from different
    settings of the 3D grid of the \texttt{L80\_N1024} box.
    Left: impact of transverse pixel size in the flux grid, for the
    default value of line-of-sight pixel width of $dz=0.078 \hMpc$.
    The default value of $dxy=0.156 \hMpc$ is enough to capture the
    effect of pressure smoothing, but the results are not fully converged.
    Right: impact of line-of-sight pixelisation in the grid.
    The default value of $dz=0.078 \hMpc$ (the best we can use on our
    larger box) does not resolve the line-of-sight structure, and this
    results in a percent difference in the large scale power, and a 
    4 \% difference on the smallest scale used in the analysis.}
    \label{fig:compare_grid}
\end{figure}

In \cref{fig:compare_grid} we compare the 3D flux power from our
default simulation box, when varying the resolution of the 3D grid.
In the left panel we vary the transverse resolution of the grid, and
show that even the default setting (blue line) is not quite converged,
with 4\% differences on the smallest scales used in the fits.
In the right panel we vary instead the line-of-sight pixelisation,
and show that the default pixelisation (red lines) caused a biased
measurement of the large-scale power of order 1\%, and a bias of
order 4\% on the smallest scales used in the fits.

\subsection{Alternative fit configurations}

\begin{figure}
    \centering
    \includegraphics[width=0.49\textwidth]{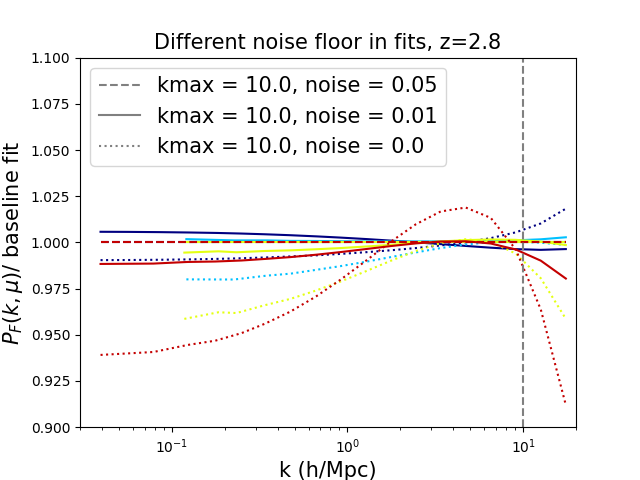}
    \includegraphics[width=0.49\textwidth]{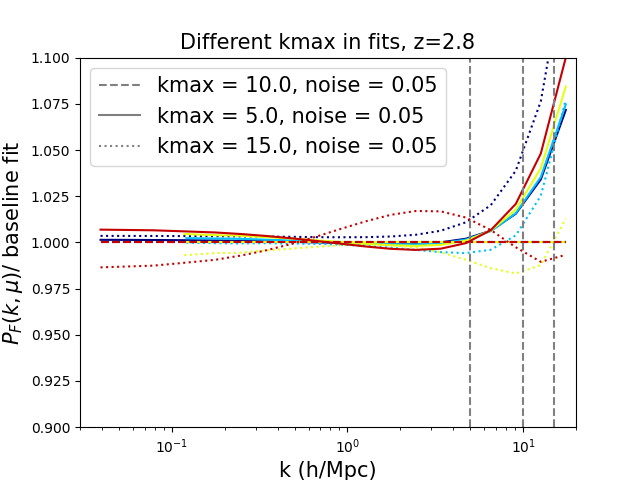}
    \caption{Ratio of best-fit models for the flux power spectra, for
    different fitting configurations on the best-fit Arinyo model.
    Left: effect of varying the noise floor $\epsilon$ in \cref{eq:weights}.
    Right: effect of varying the maximum wavenumber $k_{\rm max}$
    included in the fit.}
    \label{fig:fit_settings}
\end{figure}


As described in \cref{sec:analysis}, the default analysis of the flux
power spectrum minimises a pseudo-$\chi^2$, defined in \cref{eq:pseudo_chi2},
that weights the different band powers based on the number of Fourier modes
that contribute to a particular bin.
In order to down-weight high-k modes, that would otherwise dominate
the fit, we follow \cite{2003ApJ...585...34M,2015JCAP...12..017A} and modify
the weights by adding a \textit{noise floor} ($\epsilon$) (\cref{eq:weights}).
The left panel of \cref{fig:fit_settings} shows the impact of $\epsilon$ on the
best-fit Arinyo model. 
While the default analyses uses the same value of $\epsilon=0.05$ used in
\cite{2015JCAP...12..017A}, a noise floor five times smaller
give almost indistinguishable results (solid lines).
Ignoring the noise floor completely (dotted lines) have a noticable impact on
the best-fit large-scale biases.

In the right panel of \cref{fig:fit_settings} we study the impact of varying the
maximum wavenumber $k_{\rm max}$ used in the fits.
Decreasing (solid) or increasing (dotted) the default value of 
$k_{\rm max} =10 \ihMpc$ have only minor effects on the best fit, mostly 
limited to the modelling of pressure on the smallest scales.

\section{Results at different redshifts}
\label{app:redshift}

Most of the analyses presented in this paper use the snapshot at $z=2.8$. 
However, the \lyaf\ data usually spans a wide range of redshift, and in this appendix
we reproduce some of the key results at different redshifts.
In order to reduce the computational requirements of this comparison, we use outputs from
the small box simulation \texttt{L80\_N1024} ($80 \hMpc$, $1024^3$ particles) and the
same grid resolution than the main analysis. 
This is justified by the convergence tests presented in \cref{app:alt}.

\begin{figure}
    \centering
    \includegraphics[width=0.8\textwidth]{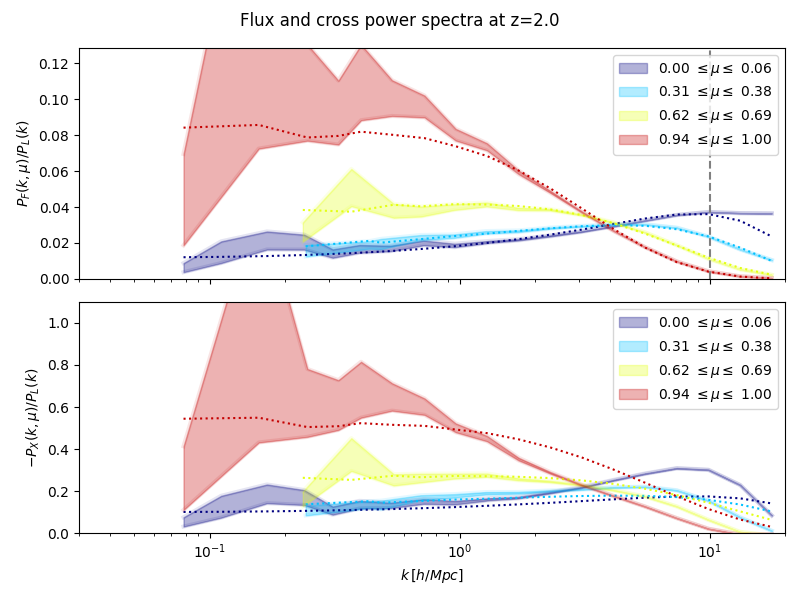}
    \caption{Measurements and models of power spectra at $z=2.0$, using the 
    \texttt{L80\_N1024} simulation with the same resolution than the one used
    in the main test.
    Top: flux power spectrum and best-fit Arinyo model, similar to \cref{fig:ari_vs_mcd}.
    Bottom: cross power-spectrum and baseline model (without free paramters) for the
    total halo catalog, similar to \cref{fig:cross}.}
    \label{fig:flux_cross_z2.0}
\end{figure}

\begin{figure}
    \centering
    \includegraphics[width=0.8\textwidth]{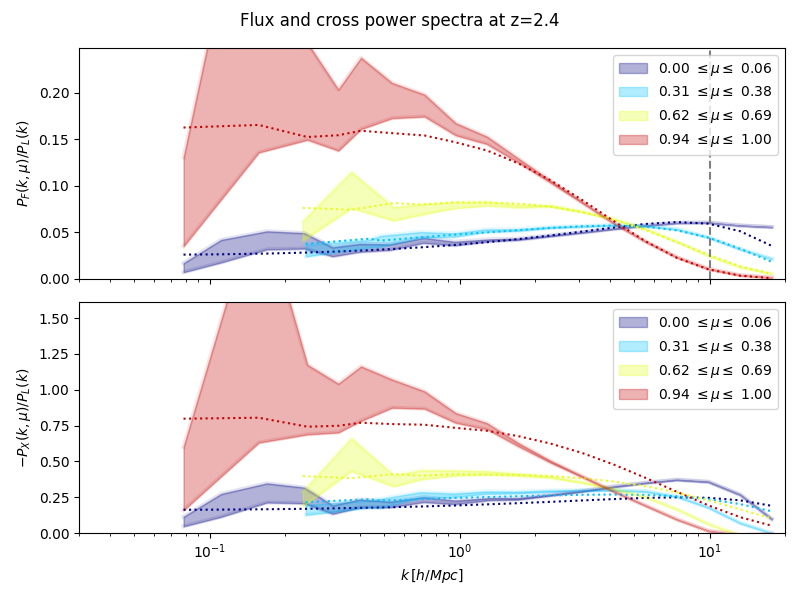}
    \caption{Same as \cref{fig:flux_cross_z2.0}, but at $z=2.4$.}
    \label{fig:flux_cross_z2.4}
\end{figure}

\begin{figure}
    \centering
    \includegraphics[width=0.8\textwidth]{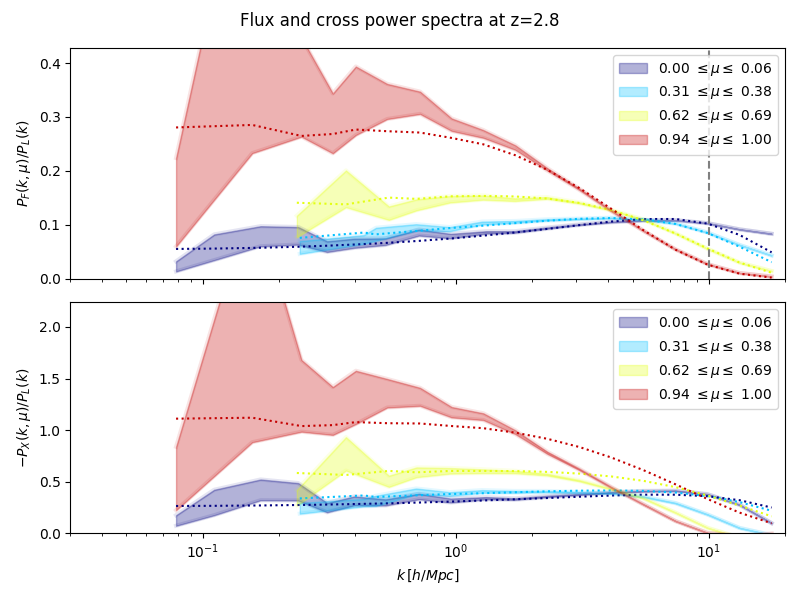}
    \caption{Same as \cref{fig:flux_cross_z2.0}, but at $z=2.8$.}
    \label{fig:flux_cross_z2.8}
\end{figure}

\begin{figure}
    \centering
    \includegraphics[width=0.8\textwidth]{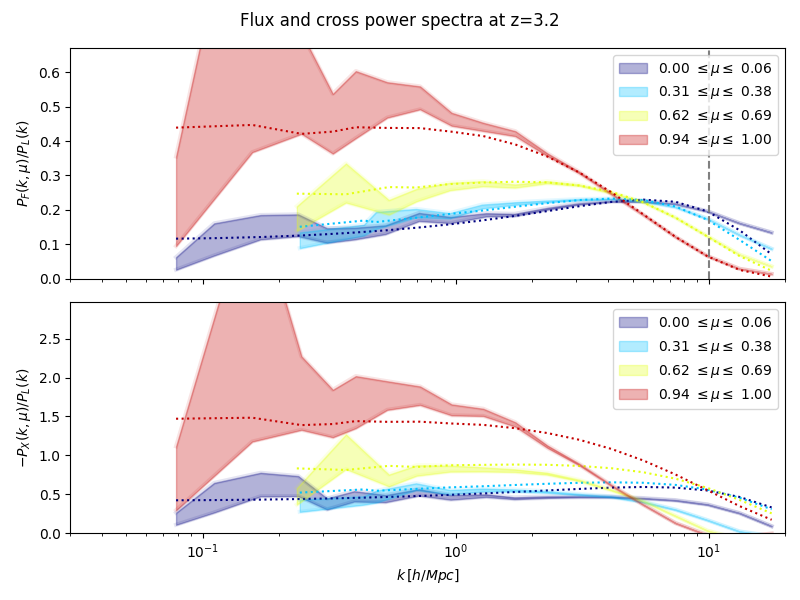}
    \caption{Same as \cref{fig:flux_cross_z2.0}, but at $z=3.2$.}
    \label{fig:flux_cross_z3.2}
\end{figure}

In \cref{fig:flux_cross_z2.0,fig:flux_cross_z2.4,fig:flux_cross_z2.8,fig:flux_cross_z3.2}
we redo the analysis of the flux power (top panels) and \lyax\ power (bottom panels) 
at redshifts $z=2.0$, $2.4$, $2.8$ and $3.2$.
At all redshifts the flux power is well described by the Arinyo model, while the 
baseline model for the \lyax\ cross-correlation (without free parameters)
is only a good description on relatively large scales.

\end{document}